\documentclass[journal=jacsat,manuscript=article]{achemso}

\usepackage[version=3]{mhchem} 
\usepackage{color}
\usepackage{adjustbox}

\author{Cecilia Cisneros}
\altaffiliation{Contributed equally to this work}
\author{Travis Thompson}
\altaffiliation{Contributed equally to this work}
\author{Noel Baluyot}
\author{Adam Smith}
\author{Enrico Tapavicza}
\email{enrico.tapavicza@csulb.edu}
\altaffiliation{Electronic supplementary information (ESI) available.}

\affiliation[California State University, Long Beach]
{Department of Chemistry and Biochemistry, California State University, Long Beach, 1250 Bellflower Boulevard, Long Beach, CA, 90840}

\title[The role of tachysterol]
  {The role of tachysterol in vitamin D photosynthesis -- A non-adiabatic molecular dynamics study}

\abbreviations{}

\begin{document}

\clearpage
\begin{abstract}
To investigate the role of tachysterol in the photophysical/photochemical regulation of vitamin D photosynthesis,  we studied its electronic absorption properties and excited state dynamics using time-dependent density functional theory (TDDFT), second-order approximate coupled cluster theory (CC2), and non-adiabatic surface hopping molecular dynamics in the gas phase.
In excellent agreement with experiments, the simulated electronic spectrum shows a broad absorption band with a remarkably higher extinction coefficient than the other vitamin D photoisomers provitamin D, lumisterol, and previtamin D.
The broad band arises from the spectral overlap of four different ground state rotamers. 
After photoexcitation, the first excited singlet state (S$_1$) decays with a lifetime of 882 fs. 
The S$_1$ dynamics is characterized by a strong twisting of the central double bond. 
In 96\% of all trajectories this is followed by unreactive relaxation to the ground state near a conical intersection. The double-bond twisting in the chemically unreactive trajectories induces a strong interconversion between the different rotamers.
In 2.3 \% of the trajectories we observed [1,5]-sigmatropic hydrogen shift forming the partly deconjugated toxisterol D1.
1.4 \% previtamin D formation is observed via hula-twist double bond isomerization.
In both reaction channels, we find a strong dependence between photoreactivity and dihedral angle conformation: hydrogen shift only occurs in cEc and cEt rotamers and double bond isomerization occurs mainly in cEc rotamers.
Hence, our study confirms the previously formed hypothesis that cEc rotamers are more prone to previtamin D formation than other isomers.
In addition, we also observe the formation of a cyclobutene-toxisterol in the hot ground state in 3 trajectories (0.7 \%).

Due to its large extinction coefficient and mostly unreactive behavior, tachysterol acts mainly as a sun shield suppressing previtamin D formation. 
Tachysterol shows stronger toxisterol formation than previtamin D and can thus be seen as the major degradation route of vitamin D.
Absorption of low energy ultraviolet light by the cEc rotamer can lead to previtamin D formation.  In addition the cyclobutene-toxisterol, which possibly reacts thermally to previtamin D, is also preferably formed at long wavelengths.
These two mechanisms are consistent with the wavelength dependent photochemistry found in experiments. 
Our study reinforces a recent hypothesis that tachysterol constitutes a source of previtamin D when only low energy ultraviolet light is available, as it is the case in winter or in the morning and evening hours of the day. 
 
\end{abstract}
\clearpage

\section{Introduction}
Vitamin D (Vita) regulates a variety of processes in our body.
Besides regulating calcium uptake and controlling bone growth,
Vita is involved in the regulation of apoptosis \cite{Dixon2005}, autoimmune diseases \cite{Wang2009}, cardiovascular diseases \cite{Wang29012008}, 
and plays a role in the natural prevention and treatment of cancer \cite{Chen2000,Tuckey2014}.
Tachysterol (Tachy) is often considered a side product in vitamin D photosynthesis, but several studies indicate its importance in the regulation of vitamin D photo production \cite{Havinga1960,MacLaughlin1982}.
Similar to other vitamin D photo isomers (DPI), also Tachy and its metabolites possess biologic activity. \cite{albright1939comparison,suda197025,Chen2000267}
For the largest part of the world population, Vita is generated by skin exposure to ultraviolet (UV) light of the sun.\cite{Holick2007} To address the widespread problem of Vita deficiency, Vita is given orally via nutrition and supplements \cite{Holick2007}. However, Vita overdoses can lead to hypercalcemia causing calcification of muscles and bones.\cite{Chen2010} 
Interestingly, Vita overdose can only be caused by oral administration, but it has never been observed by extended sun exposure.
This is due to a self-regulation mechanism based on intrinsic photophysical and photochemical properties of the involved DPI, that leads to quenching of previtamin D (Pre) production under prolonged sun irradiation preventing the overproduction of Vita. The regulation responds within seconds and is not regulated by pigment formation\cite{Holick1981}.
However, the regulation of Pre production only functions properly under irradiation of light with the specific wavelength distribution of the solar spectrum, while it fails under monochromatic irradiation \cite{Havinga1960}. 
To quantitatively explain this phenomenon, it is necessary to understand the action spectrum of Vita production, i.e., the efficiency of photoinduced vitamin D production as a function of  the wavelength of monochromatically irradiated UV light. However, since the regulative mechanism only works under irradiation of the spectrum of a black body emitter at the temperature of the sun, it is questionable 
to which extent the monochromatically derived action spectrum is capable to explain the photoequilibrium under the solar spectrum.
At present, there is an ongoing dispute about the validity of several measured action spectra in the literature\cite{Norval2010}, and a recent study has found that the action spectra vary between previously irradiated and unexposed skin samples\cite{vanDijk2016}.
To date, the exact regulative mechanism of Vita photo production is still not well understood and no accurate quantitative predictions can be made.\cite{vanDijk2016}
Just recently it has been found that previtamin D can be generated to substantial amounts from tachysterol by irradiation at wavelengths above 315-340~nm\cite{andreo2015generation}. This is in stark contrast to the common belief that vitamin D cannot be produced at wavelengths longer than 320 nm. 
With respect to the vitamin D deficiency problem, this is an important finding, because this means that even in winter at northern latitudes or in the morning and evening hours of the day, where high energetic UV radiation is not available, previtamin D could be synthesized by sun exposure. 
This indicates that Tachy can act a as previtamin D reservoir which is charged under conditions where high energetic UV radiation is present, but can be later used as previtamin D source at longer UV wavelengths.
We want to test this hypothesis by molecular dynamics simulation.

\begin{figure}[h]
  \includegraphics[scale=0.5]{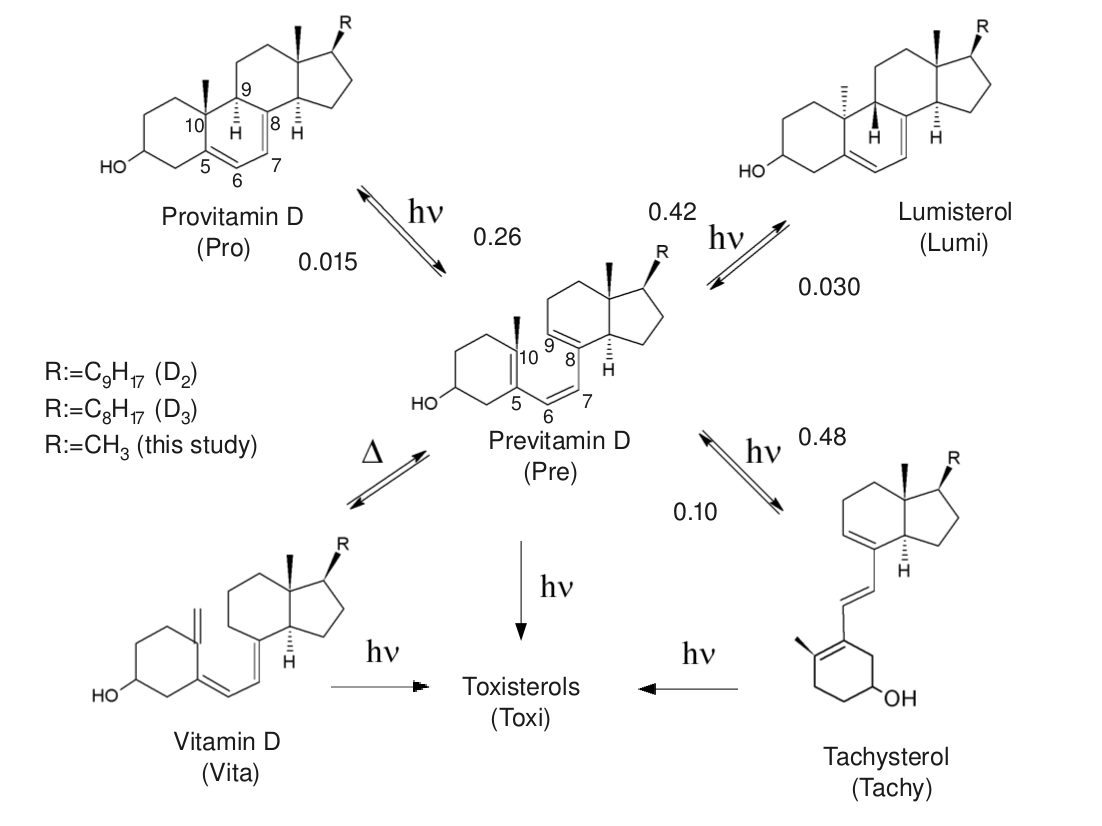}
  \caption{\label{scheme:pre_hub} (Photo)chemical reactions involved in the formation of vitamin D. Rate constants, measured in ether at 0$^{\circ}$ C \cite{Havinga1973}, are given on the arrows for some reactions.}
\end{figure}

Photochemical vitamin D production starts with provitamin D (Pro) ring-opening, forming previtamin D (Pre), which then isomerizes thermally to vitamin D via [1,7]-hydrogen shift (Fig. \ref{scheme:pre_hub}).
Pre itself plays the central role, as it reacts to four distinct main Vita photo isomers (DPI) Pro, Lumi, and Tachy and toxisterols (Toxi).
While the provitamin D ring-opening and its analogous reaction in cyclohexadiene has been studied extensively, both theoretically\cite{Tapavicza2011,Schalk2016,Snyder2016,lei2016photo} and experimentally\cite{Fuss1996,Fuss2000,Anderson1999,Tang2011,Arruda2013,Schalk2016,Smith2016},
 little research has been done to assess the influence of tachysterol on photochemical vitamin D production.  In particular, its excited state dynamics has neither been investigated by experiments nor by simulations.
 Tachysterol is mainly formed by UV excitation of tZt conformers of previtamin D \cite{Tapavicza2011} (Fig. \ref{scheme:pre_hub}). Experimentally, a quantum yield of 0.29 has been found on the red side of Pre's absorption spectrum
 (302.5 nm), whereas an even higher quantum yield of of 0.41 was found at 254 nm on the blue side close to the absorption maximum.\cite{Jacobs1981} Cis-trans isomerization of the central double bond occurs via the hula-twist mechanism and is thought to be the cause of this isomerization, which has been confirmed experimentally \cite{Mueller1998} and by non-adiabatic molecular dynamics simulations,\cite{Tapavicza2011} but also questioned at low temperature.\cite{redwood2013photoisomerization} 
Under continuous solar irradiation, a quasi-stationary photoequilibrium is adopted and the Pre concentrations do not further increase \cite{MacLaughlin1982}.
The self-regulation mechanism is most likely a consequence of an evolutionary fine tuning of several factors.\cite{Holick2011} 
First, the four major DPI possess distinct absorption spectra, with tachysterol having the largest extinction, covering the absorption bands of all other DPI.
A possible reason preventing the quantitative modeling of vitamin D photo kinetics could be the neglect of the conformational dependent photochemistry of the seco-steroids Pre and Tachy, which is difficult to assess experimentally.
Ab initio molecular dynamics can give structural information and its relationship to photochemical properties.
 Secondly, interconversion rates and quantum yields of the involved photo reactions differ. Furthermore, most of the photochemical reactions are highly wavelength dependent and controlled by the conformational equilibrium of different rotamers of the secosteroid molecules.\cite{Tapavicza2011} 
At present it is unclear if the interconversion rates between the different rotational isomers play a role in the photochemical regulation.
Another factor in the quenching of Vita production is the degradation of DPIs to toxisterols.\cite{terenetskaya2008limitations}
Furthermore, since the reaction naturally takes place in the cellular membrane of the epidermis, also the steric interactions between DPI and the biological membrane might play a role.\cite{Tian1993,Tian1999} The membrane is thought to enhance the isomerization from Pre to Vita, by enforcing the cZc conformation necessary to allow for the [1,7]-sigmatropic hydrogen shift. 
To explain quantitatively the kinetics of the quasi stationary photoequilibrium depicted in Fig. \ref{scheme:pre_hub}, it is necessary to understand the kinetics of the individual photochemical reactions and also the isomerization rates between different rotational isomers of the DPI. To shed light on the role of tachysterol in the photophysical self-regulation we study the photodynamics of this compound using non-adiabatic molecular dynamics simulations based on time-dependent density functional theory (TDDFT).


To this end, we will first assess the equilibrium of ground state rotamers of tachysterol. 
We then study the influence of rotamer conformation on the absorption spectrum and on its photochemical reactivity. 
The results of our simulation are put in context with other computational and experimental findings on the vitamin D photoequilibrium.


\section{Methods and computational details}
To study the photochemical reactivity of the macroscopic ensemble of a compound and to assess its relationship to conformation and excitation wavelength,
 it is necessary to generate a Boltzmann ensemble of the possible structures with their correct statistical weight. To generate this ensemble, we apply Born-Oppenheimer molecular dynamics (BOMD) using ab initio nuclear gradients. Once the ensemble is generated, the overall absorption spectrum is calculated as an average of the individual absorption spectra of snapshot structures from the BOMD simulations. The photochemical reactivity and characteristic product distribution can be studied using non-adiabatic molecular dynamics in the photoexcited electronic state.
In the following we will outline this procedure and give specific details to the calculations.

\subsection{Ground state molecular dynamics}
The distributions of ground state conformers of the rigid closed-ring steroids provitamin D and lumisterol were sampled by ab initio BOMD\cite{Elliott2000}.
To effectively sample the ground state structures of the conformationally flexible open-ring secosteroids Tachy and Pre, we employed ab initio replica exchange molecular dynamics (REMD) \cite{Sugita1999}.
In REMD, several BOMD trajectories at different temperatures are computed in parallel. After a certain number of simulation time steps, 
a switching probability between two structures $i$ and $j$ for two temperatures $T_i$ and $T_j$ is calculated.
The probability is based on the Boltzmann weight of the two structures at the different temperatures:\cite{Sugita1999}
\begin{equation}
P(i\rightarrow j)=\min\left(1,\frac{\exp\left(\frac{E_j}{kT_i}-\frac{E_i}{kT_j}\right)}{\exp\left(\frac{E_j}{kT_j}-\frac{E_i}{kT_i}\right)}     \right)
\end{equation}
This method has been shown to accelerate sampling of systems with several separated minima.

In both, conventional BOMD and REMD simulations, a simulation time step of 50 au was used and the temperature was controlled by a Nos\'e-Hoover thermostat \cite{Nose1984,Hoover1985} with a characteristic response time of 500 au.
REMD simulations were performed at four different temperatures, i.e., 300 K, 600 K, 900 K, and 1200 K. Geometry switches between the simulations at different temperatures were carried out every 200 MD steps. 
Total simulation time of the BOMD of Pro and Lumi amounts to approximately 24 ps; for Tachy and Pre REMD was carried out about a total of approximately 125 ps and 87 ps, respectively.
In all ground state MD simulations the PBE functional\cite{Perdew1996} and the SVP basis set \cite{SVP} was used. Calculations were accelerated using the resolution of identity approximation \cite{Eichkorn1997}.

\subsection{Electronic absorption spectra}
Absorption spectra of the different vitamin D isomers (DPI) were calculated using snapshot structures generated by BOMD and REMD, as previously described \cite{Epstein2013, Tapavicza2013}. 
For the closed ring DPI Pro and Lumi 217 and 295 structures, respectively, were randomly chosen from the ground state MD trajectory.
For the flexible open-ring secosteroids Pre and Tachy, 972 and 1349 structures were used.
Based on these single point structures, we computed the absorption spectra using TDDFT in combination with the hybrid exchange-correlation functional PBE0 (TDPBE0) and second-order approximate coupled cluster singles and doubles (CC2) \cite{Christiansen95,Hattig02} within the resolution of identity approximation \cite{Hattig03}.
In both cases, the lowest two excited states were calculated.
To convert the oscillator strengths to the molar extinction coefficients, the individual spectra were broadened using a Gaussian line shape with a full width at half-maximum of 0.1 eV. The broadened spectra were then averaged to obtain the absorption spectrum of the macroscopic ensemble.\cite{Epstein2013}
Although this treatment neglects the quantum nature of the nuclear vibrations\cite{Tapavicza2016}, the influence of different conformations will be visible in the spectrum.

\subsection{Non-adiabatic dynamics}
Tully's fewest switches surface hopping Surface hopping method \cite{Tully1990} has been successfully applied to simulate photochemical reactions in a variety of systems \cite{Doltsinis2002,Barbatti2007,Tapavicza2009,barbatti2011nonadiabatic,Nelson2011,Mitric2011a,Tapavicza2011,yu2014trajectory,Tapavicza2013,curchod2013trajectory}.
Despite several shortcomings due to neglect of parts of the quantum nature of the nuclei, the method has the advantage to be computational efficient and gives qualitatively accurate results\cite{Granucci2007}.
The method is based on the Born-Oppenheimer (BO) expansion to describe the time-dependent wavefunction,\cite{Domcke2004conical} 

\begin{equation}
\label{eq:BOwavefunction}
|\Psi(t|\mathbf{R})\rangle = \sum_{n}c_{n}(t|\mathbf{R})|\Phi_{n}(\mathbf{R})\rangle,
\end{equation} 
where $c_{n}(t|\mathbf{R})$ are the state amplitudes of the BO states $\Phi_{n}(\mathbf{R})$, both depending on the nuclear positions $\mathbf{R}$.
Inserting Eq.~(\ref{eq:BOwavefunction}), into the time-dependent Schr{\"o}dinger equation yields a coupled differential equation for the time-dependent amplitude vector $\mathbf{c}$

\begin{equation}
\label{eq:amplitudevector}
i\dot{\mathbf{c}} = (\mathbf{H} - i{\mathbf{Q}})\mathbf{c} .
\end{equation}
Here ${H}_{mn} = E_{mn}\delta_{mn}$ are the Born-Oppenheimer energies and $Q_{mn}$ is the first-order derivative coupling between Born-Oppenheimer states $m$ and $n$,
\begin{align}\label{eq:NAC}
Q_{mn}=\langle\Phi_m|\frac{\partial}{\partial t}\Phi_n\rangle=\dot{\mathbf{R}}\cdot\boldsymbol{\tau}_{mn}.
\end{align}
Here, ${\boldsymbol{\tau}}_{mn}(\mathbf{R})$ are the Cartesian first-order non-adiabatic derivative coupling vectors, 
\begin{equation}
\label{eq:couplingvector}
\boldsymbol{\tau}_{mn}(\mathbf{R}) = \langle\Phi_{m}(\mathbf{R})|\frac{\partial}{\partial{\mathbf{R}}}|\Phi_{n}(\mathbf{R})\rangle.
\end{equation}
At each MD step the instantaneous state amplitudes ${c}_{n}$ are computed by integration of Eq. \ref{eq:amplitudevector}. 
The transition probability between states $m$ and $n$ is given by \cite{Tully1990}
\begin{equation}
\label{eq:tully}
g_{mn} = \frac{\Delta{t}\sum_{l \neq n}[2\operatorname{Im}({c}^{*}_{n}{H}_{nl}c_{l} + {c}^{*}_{n}{V}_{ext,nl}c_{l}) - 2\operatorname{Re}({c}^{*}_{n}\dot{\mathbf{R}}\cdot{\mathbf{\tau}}_{nl}c_{l})]}{|{c}_{m}|^2}.
\end{equation}
If $g_{mn}$ is larger than a random number between 0 and 1, then the trajectory is switched from state $m$ to state $n$. 
In case of a switch, the momenta of the nuclei are scaled along the non-adiabatic coupling vector to conserve the total energy.\cite{herman2949,herman2771,Vincent2016} 
Between hops, the individual trajectories evolve on a single BO potential energy surface, according to Newton's equations of motion.

The surface hopping algorithm requires the computation of excited state potential energy surfaces, nuclear forces, non-adiabatic derivative couplings\cite{Domcke2004conical} that mediate transitions between the excited state potential energy surfaces.
Excited state energies, analytical nuclear forces\cite{Furche2002} and non-adiabatic couplings between electronic states are available from time-dependent linear response theory \cite{Chernyak2000,Baer2002,Tavernelli2009,Tavernelli2009a,Send2010,Ou2015_nac_between_excited_states}.
In particular the local basis set implementation of time-dependent density functional theory surface hopping (TDDFT-SH) \cite{Tapavicza2007,Tapavicza2011,Tapavicza2013} offers a computationally efficient and accurate description as it allows employment of hybrid exchange-correlation functionals, necessary to achieve accurate description of charge transfer excitations in medium sized molecules \cite{Send2011}.
However, employing TDDFT to study photodynamics, one has to pay special attention to several potential problems associated with the commonly used exchange-correlation functionals. Several shortcomings have been identified to be a concern in TDDFT calculations. Problematic for photodynamics are the under estimation of charge-transfer excitations\cite{Dreuw2003}, erroneous description of double-excitations and conical intersections\cite{Levine2006}. The mentioned problems can be controlled by the usage of selected functionals and careful assessment of the accuracy by comparison with more accurate excited state electronic structure methods. In particular the usage of the hybrid exchange-correlation functional PBE0 has been shown to achieve good accuracy with respect to charge-transfer excitations in medium-sized molecules\cite{Send2011}. The importance of double excitations can be assessed by comparison with second-order approximate coupled cluster singles and doubles (CC2) theory calculations. If the double amplitudes are below 10 \% a good accuracy is suggested \cite{Christiansen1996,Christiansen1996a}.
The problem of erroneous dimensionality and instabilities at conical intersections\cite{Levine2006,Tapavicza2008} can be minimized by the usage of the Tamm-Dancoff approximation to TDDFT\cite{Hirata1999}. Previous calculations have shown that globally intersections exhibit correct dimensionality and only a local area of instabilities exist.\cite{Tapavicza2008}.
The localized basis set implementation of the TDDFT-SH method has been described in detail previously \cite{Tapavicza2011,Tapavicza2013}. 

For tachysterol, a total of 442 REMD snapshot structures and their instantaneous ground state velocities were used as starting conditions for TDDFT-SH simulations. 
Trajectories were started in the first excited singlet state ($S_1$) and propagated for over 2 ps.
The total energy was kept constant (NVE ensemble), assuming no energy dissipation to the environment. A time step of 50 au was used to propagate the classical nuclear degrees of freedom.

\subsection{Additional computational details}
Since the side chain (R in Fig. \ref{scheme:pre_hub}) is not expected to have large influence on the photochemistry, we replaced it by a methyl group in all DPI. All calculations were carried out with the TURBOMOLE quantum chemistry package \cite{TURBOMOLE}.
Ground state density functional theory (DFT) calculations were performed using the dscf module \cite{Haeser1989}.
Excited state calculations employ the time-dependent density functional theory implementation \cite{Furche2002}.
Molecular Dynamics calculations were carried out using the Verlet algorithm, as implemented in the frog module \cite{Elliott2000}.
Non-adiabatic dynamics were carried out using the surface hopping (TDDFT-SH) method \cite{Tapavicza2011,Tapavicza2013}. 
For excited state dynamics, TDDFT was used within the Tamm-Dancoff approximation \cite{Hirata1999}, whereas for spectra calculations we employ the full linear response equations \cite{Furche2002}, which is expected to give more reliable oscillator strengths since it obeys the Thomas-Reiche-Kuhn sum rule\cite{Furche2001}.

\section{Results and discussion}
\subsection{Ground state equilibrium of tachysterol}
\begin{figure}
\includegraphics[scale=0.4]{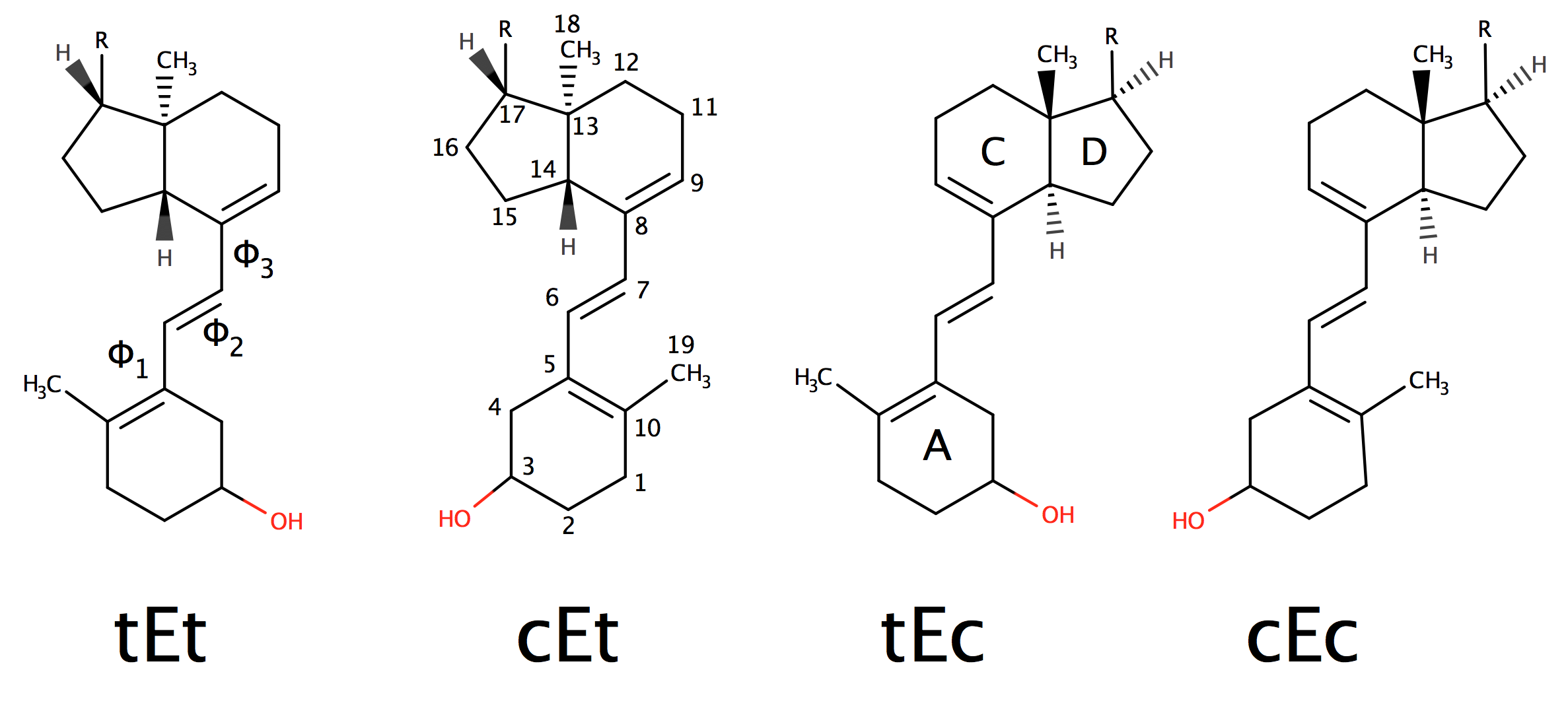}
\caption{\label{rotamers} Tachysterol rotamers. Dihedral angle $\phi_1$ is defined by carbons C10-C5-C6-C7, $\phi_2$ is defined by carbons C5-C6-C7-C8, and $\phi_3$ is defined by carbons C6-C7-C8-C9. Nomenclature according to Havinga\cite{Havinga1973}. Rotamers were classified according to their dihedral angle conformation: tEt: $\phi_1=[90^{\circ};180^{\circ}]$ or $[-180^{\circ};-90^{\circ}]$ and $\phi_3=[90^{\circ};180^{\circ}]$ or $[-180^{\circ};-90^{\circ}]$; cEt: $\phi_1=[-90^{\circ};90^{\circ}]$ and $\phi_3=[90^{\circ};180^{\circ}]$ or $[-180^{\circ};-90^{\circ}]$; tEc: $\phi_1=[90^{\circ};180^{\circ}]$ or $[-180^{\circ};-90^{\circ}]$ and $\phi_3=[-90^{\circ};90^{\circ}]$; cEc: $\phi_1=[-90^{\circ};90^{\circ}]$ and $\phi_3=[-90^{\circ};90^{\circ}]$.}
\end{figure}

We analyzed 87907 structures of the 300 K ensemble obtained from the REMD simulation and classified them according to their dihedral angle conformation $\phi_1$ and $\phi_3$ into four different rotamers: tEt, cEt, tEc, and cEc (Fig. \ref{rotamers}). As shown in the histogram (Fig. \ref{histogram}) and Table \ref{rotamer_dist_2}, the tEc rotamers are present with the highest statistical weight of 46 \%, followed by tEt rotamers (26 \%). Rotamers cEc and cEt represent 15 \% and 13 \% of the structures, respectively. Regarding the major and the minor rotamer, our distribution is consistent with the equilibrium previously found using static molecular mechanics (MM) calculations \cite{Dmitrenko1997} (Table \ref{rotamer_dist_2}), which also predicts tEc as the main rotamer (66 \%) and cEt as the rotamer with least statistical weight (6 \%). 
However, our simulations predict a higher percentage of tEt rotamers than cEc rotamers, which leads to a qualitatively different order than the MM results (tEt: 13 \%; cEc: 18 \%).


\subsection{Electronic absorption spectra of tachysterol and vitamin D photoisomers}
We optimized the ground state structures of the different Tachy rotamers (coordinates are given in the ESI) and calculated the $S_{1}\leftarrow S_0$ excitation energies (Table \ref{opt_exci}).
Comparing the $S_{1}\leftarrow S_0$ excitation energy ($\omega_1$) of the tEc rotamer, which has the highest statistical weight, with the experimental $\lambda_{\text{max}}$ of 4.428 eV, CC2/TZVP gives the best result with 4.53 eV. In general, CC2/TZVP values range from 4.39-4.73eV and show smallest deviations from the experiment. CC2/SVP leads to an overestimation of 0.1-0.2 eV with respect to CC2/TZVP. Compared to the experimental $\lambda_{\text{max}}$, TDPBE0/SVP and TDPBE0/TZVP underestimate $\omega_1$ of the cEt rotamer by 0.3 and 0.4 eV, respectively.  TDPBE0/SVP values range from 3.98-4.26 eV and TDPBE0/TZVP all about 0.1 eV lower. All methods predict the S$_1$-S$_2$ gap to be 1$\pm$0.1 eV independent of the basis set.
Both methods also predict S$_1$ oscillator strengths to be much larger than S$_2$ oscillator strengths. 
CC2 predicts generally larger oscillator strengths than TDPBE0 (Fig. 1, ESI).
In all cases, CC2 T2-amplitudes are below 10\%, which indicates minor importance of double excitations for S$_1$ and S$_2$\cite{Christiansen1996,Christiansen1996a}.
$S_{1}\leftarrow S_0$ excitation energies for 1349 Tachy structures of the Boltzmann ensemble  range from 3.50-5.38 eV for TDPBE0 and from 4.11-6.29 eV for CC2. Thus, CC2 values have a larger spread of 2.2 eV than the TDPBE0 values (1.9 eV). 
On average, CC2 calculations predict 5.8 \% double contribution in the first excited state 
with a maximum value of 6.5 \% and a minimum of 4.8 \%. This is below the 10 \% threshold, indicating that that double excitations are not of main importance and thus TDDFT excitation energies can be trusted\cite{Christiansen1996,Christiansen1996a}.
For both TDPBE0 and CC2, we find a similar dependency of the excitation energy and oscillator strengths on the dihedral angle conformation, with cEc and cEt conformers exhibiting lower average excitation energies and tEt and tEc exhibiting excitation energies on the blue side of the spectrum (Fig. \ref{specosci}). tEt rotamers exhibit highest oscillator strengths, followed by tEc. cEc and cEt rotamers have lowest oscillator strengths. 
Averaging the spectra of the individual snapshot structures and broadening with a Gaussian line width leads to the broad absorption spectrum of tachysterol in good agreement with the experimental spectrum (Fig. \ref{spec_rotamers}).
The shapes of the resulting spectral bands obtained from TDDFT and CC2 agree well with each other, but the maximum of the experimental extinction is slightly smaller than in both calculated spectra. 
With respect to the experimental position of the peak maximum (280 nm, 4.428 eV), TDPBE0 underestimates the position of the peak maximum by 0.350 eV and CC2 overestimates it by 0.251 eV (Table \ref{spec_compare}). 
In regard to the CC2/SVP and CC2/TZVP excitation energies for the optimized structure (Table \ref{opt_exci}), it is expected to obtain slightly better agreement (within 0.15 eV) with experiment employing the TZVP basis in CC2 calculations. For TDPBE0, usage of the TZVP is expected to lead to a slightly larger underestimation.  
Overall, for both TDPBE0 and CC2, the agreement with the experiment is similar to the previously reported accuracy\cite{Send2011,Ramakrishnan2015}.

A closer analysis of the relationship between dihedral angle conformation and absorption band reveals that
the broad absorption band of Tachy is caused by the bands of the different rotamers that are overlapping to large parts, but still vary in the position of their maxima (Fig. \ref{spec_rotamers}). 
The main contribution of the absorption stems from tEc and tEt rotamers, not only because they exhibit the highest oscillator strengths but also because they have the highest statistical weight of structures. cEt and cEc have lower contributions to the total absorption because of their low oscillator strengths and low statistical weight.
In general, cEc and cEt absorb at the red side of the spectrum, whereas tEt and tEc absorb at the blue side of the spectrum.

Our results are somewhat different from the assignment of Saltiel et al.\cite{Saltiel2003}, who assigned the shoulder at the onset of the Tachy absorption spectrum to the cEc rotamers. Our calculation, however, shows that the shoulder is mainly caused by the {tEc rotamers}, but also contains a small contribution of the cEc rotamers. At the very long wavelength side (3.8-4.0 eV in Fig. \ref{spec_rotamers}), cEc absorption becomes more dominant, which is an important result that explains the wavelength dependent photochemistry,
in particular the Pre formation at longer wavelengths, as we will see in the next section.

Evaluating the positions of the spectra of the individual rotamers relative to the radiation influx of the sun (Fig. \ref{spec_rotamers}), we see that cEt and cEc rotamers exhibit larger overlap with the global spectral flux\cite{green1974middle} at the long wavelength side of the Tachy absorption spectrum than tEc and tEt rotamers.
Relative to the total absorption of sun light of Tachy, we find increased contributions of cEt and cEc rotamers than expected from their statistical contribution (Table \ref{opt_exci}).
 The latter two rotamers are therefore expected to have larger importance in the photochemistry under solar irradiation.
In particular the cEc rotamer, that contributes 15.2 \% of the structures, has a much larger contribution (32.8~\%) to the total spectral overlap of Tachy and the global flux (Table \ref{rotamer_dist_2}). The weight becomes even larger at increasing zenith angle, where the maximum of the irradiation density is shifted to longer wavelengths.

Comparing the absorption spectrum of Tachy with the spectra computed for the other main DPI (Fig. \ref{spec_comparison}), Lumi, Pro, and Pre, we note an increased extinction coefficient, which is consistent with the experiment.  
However, for both, TDPBE0 and CC2, compared to the experiment the spectrum of Tachy is red-shifted relative to the other DPI, and does not completely cover the spectra of the other DPI on the blue side of the spectrum. 
The red shift is smaller for CC2 than for TDPBE0. At the high energy side of the spectrum, differences between experimental and calculated spectra are most likely caused by the neglect of excited states above S$_2$, which was made in our calculations. However, since the overlap of the spectrum with global flux is negligible in this region, the deviations have no influence when we consider the photochemistry under natural solar irradiation.


\subsection{Excited state dynamics}
\begin{figure}
\includegraphics[scale=0.55]{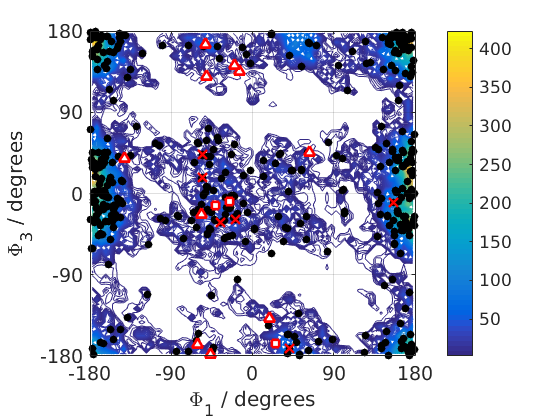}
\includegraphics[scale=0.55]{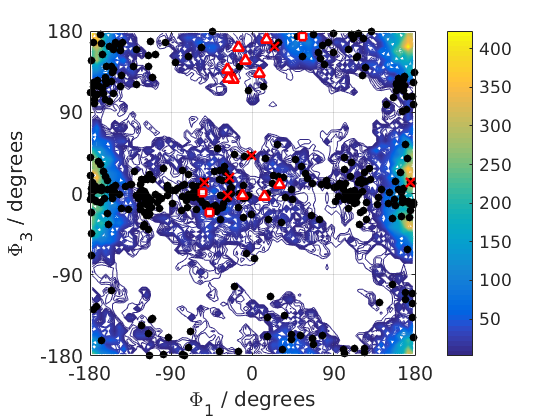}
\caption{\label{histogram} Distribution of rotamers as a function of the dihedral angles $\phi_1$ and $\phi_3$ in the ground state equilibrium of tachysterol obtained from REMD at 300 K is indicated by the contours left and right. 
Number of structures per 10$\times$10 deg$^2$ are indicated by the color code.
Left: Position of TDDFT-SH starting structures are indicated by the symbols.
Right: Position of TDDFT-SH structures at the time of the surface hop are indicated by the symbols. 
Black point: unreactive (product Tachy); cross: trans-cis isomerization (product Pre), triangle: hydrogen transfer (Toxisterol D1); square: thermal cyclobutene (CB) formation (product: CB-Toxisterol). 
}
\end{figure}

\begin{figure}
\includegraphics[scale=0.45]{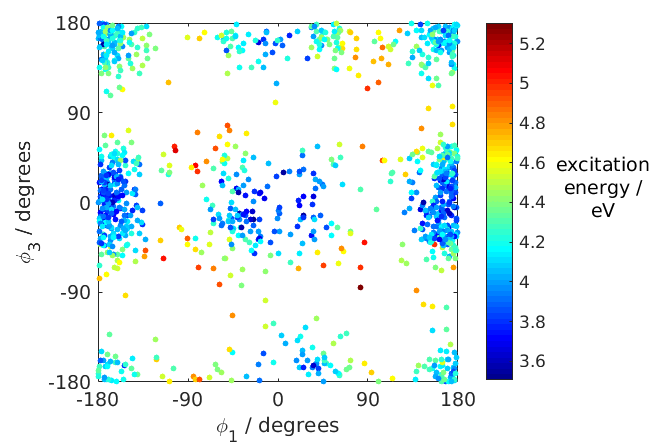}
\includegraphics[scale=0.45]{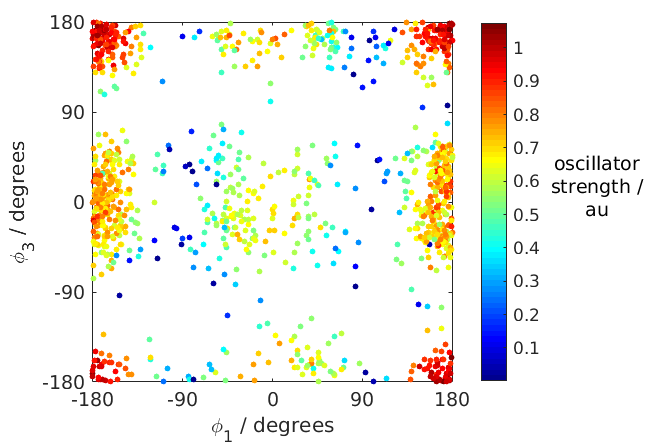}
\caption{\label{specosci} TDPBE0 S$_1\leftarrow$S$_0$ excitation energy (left) and oscillator strength (right) as a function the dihedral angles $\phi_1$ and $\phi_3$, as defined in Fig. \ref{rotamers}. For CC2 results see Fig.~1 in the ESI.}
\end{figure}

\begin{figure}
\includegraphics[scale=0.2]{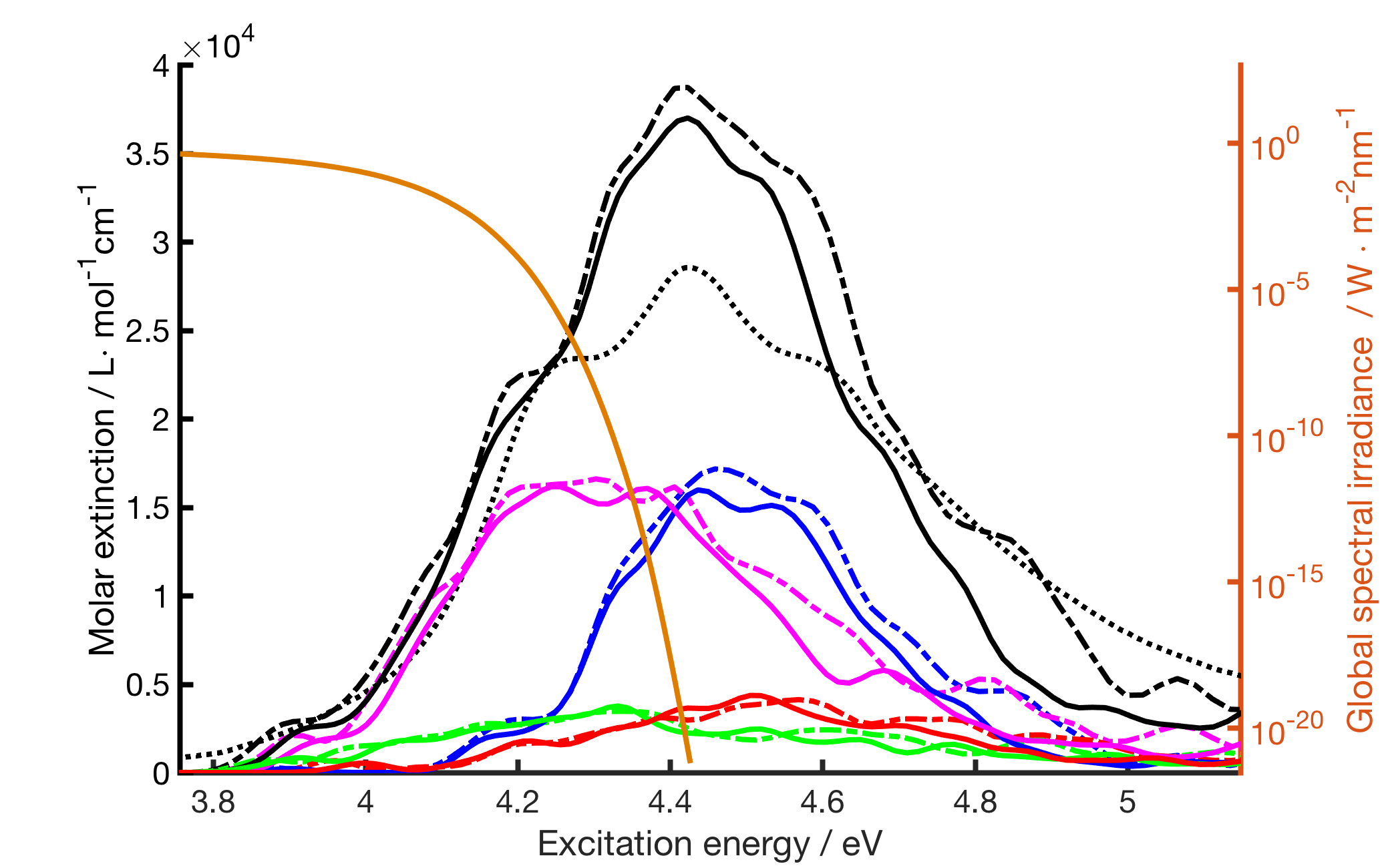}
\caption{\label{spec_rotamers} Tachysterol absorption spectrum for the different rotamers computed by TDPBE0 (solid) and CC2 (dashed). tEt: blue, tEc: magenta, cEt: red, cEc: green. Experimental spectrum (dotted) measured in ether \cite{MacLaughlin1982}. TDPBE0 and CC2 absorption bands were shifted by +0.350 eV and -0.251 eV, respectively to match the position of the peak maximum of the experimental spectrum. As example, the global irradiation energy density for a zenith angle of 30$^{\circ}$ is shown (brown)\cite{green1974middle}. These conditions approximately correspond to the irradiation in Berlin on July 8 at noon \cite{solar_position}.
}
\end{figure}

\begin{figure}
\includegraphics[scale=0.2]{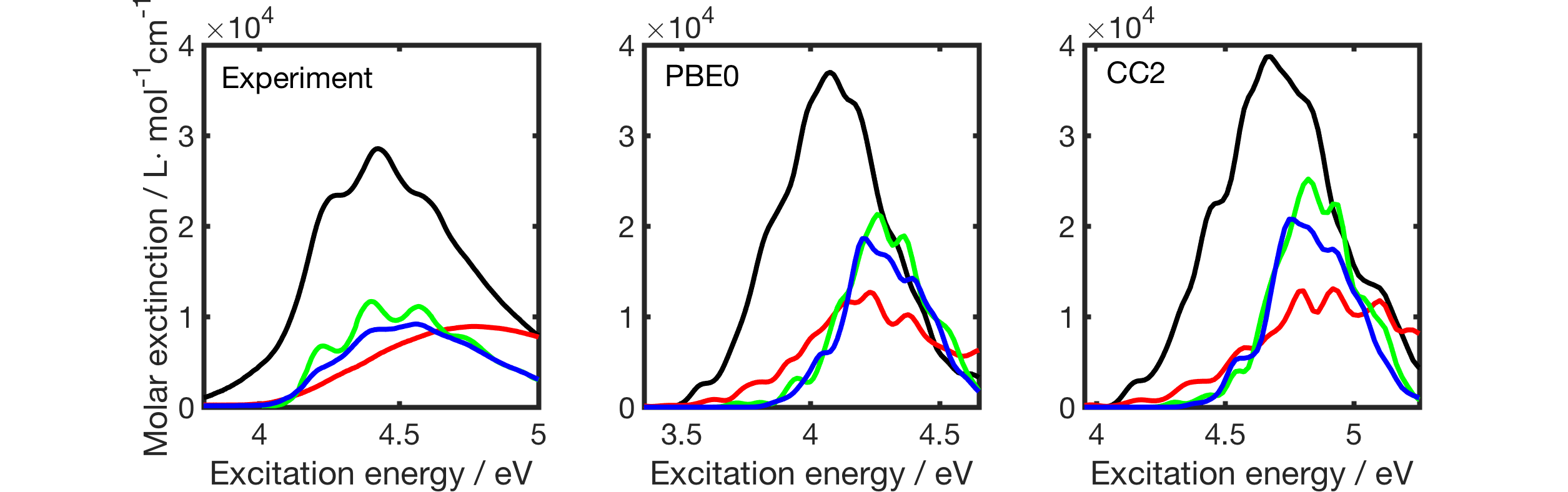}
\caption{\label{spec_comparison} {Comparison between the experimental absorption spectra of vitamin D photoisomers measured in ether \cite{MacLaughlin1982} (left) and the spectra calculated by TDPBE0 (middle) and CC2 (right). Tachysterol: black, provitamin D: green, previtamin D: red, lumisterol: blue.}}
\end{figure}

\begin{figure}
\includegraphics[scale=0.16]{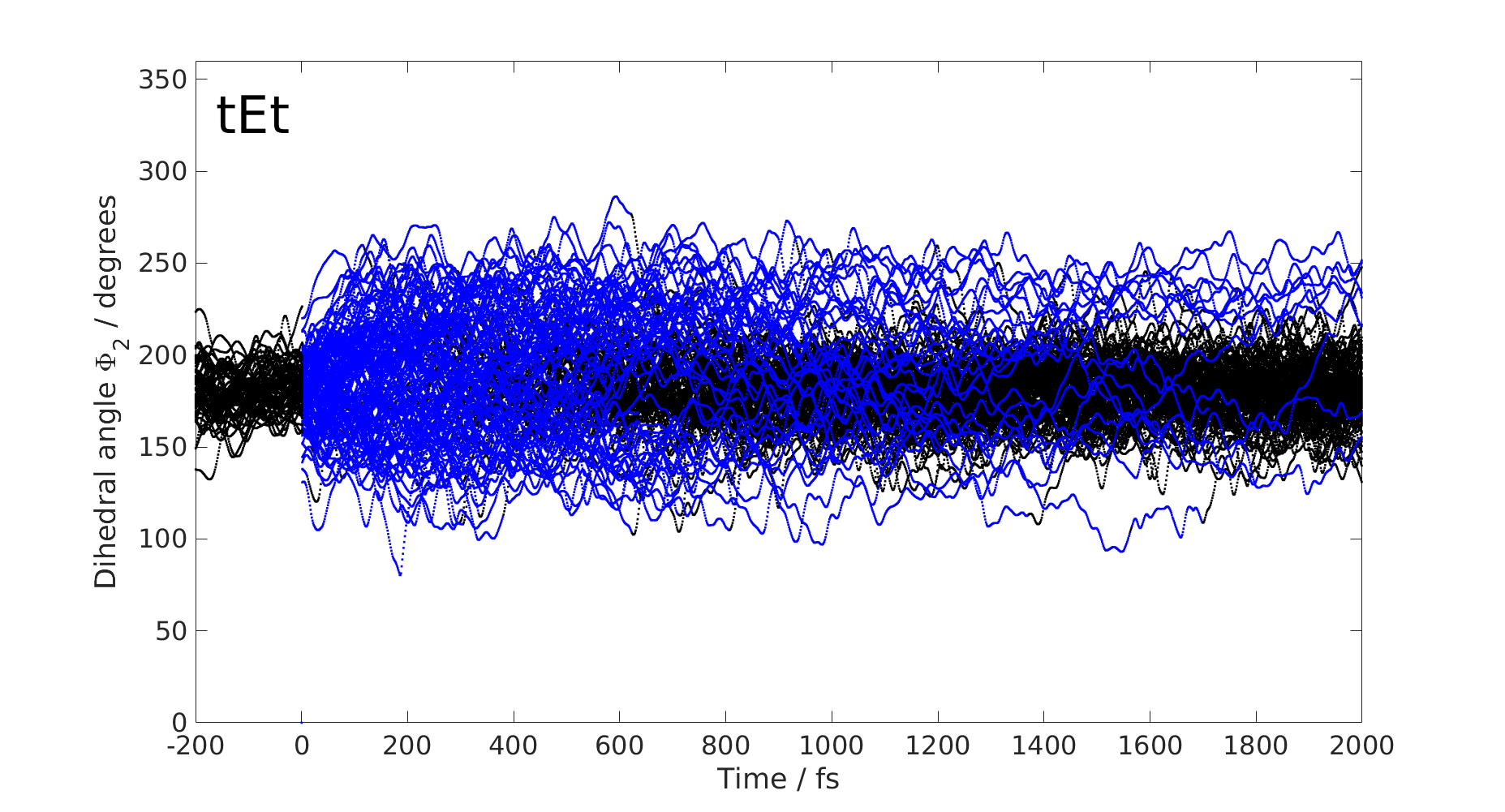}
\includegraphics[scale=0.16]{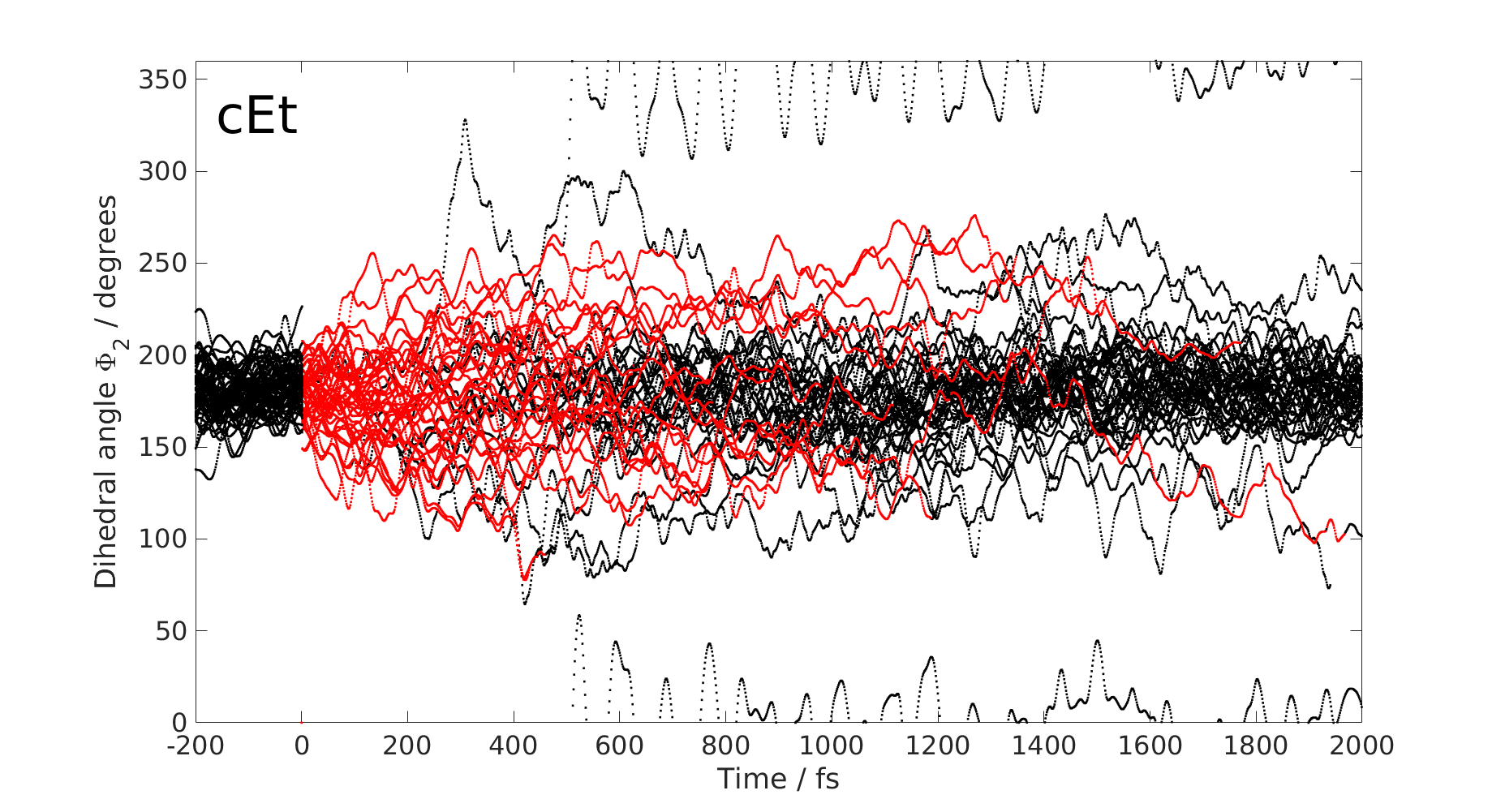}
\includegraphics[scale=0.16]{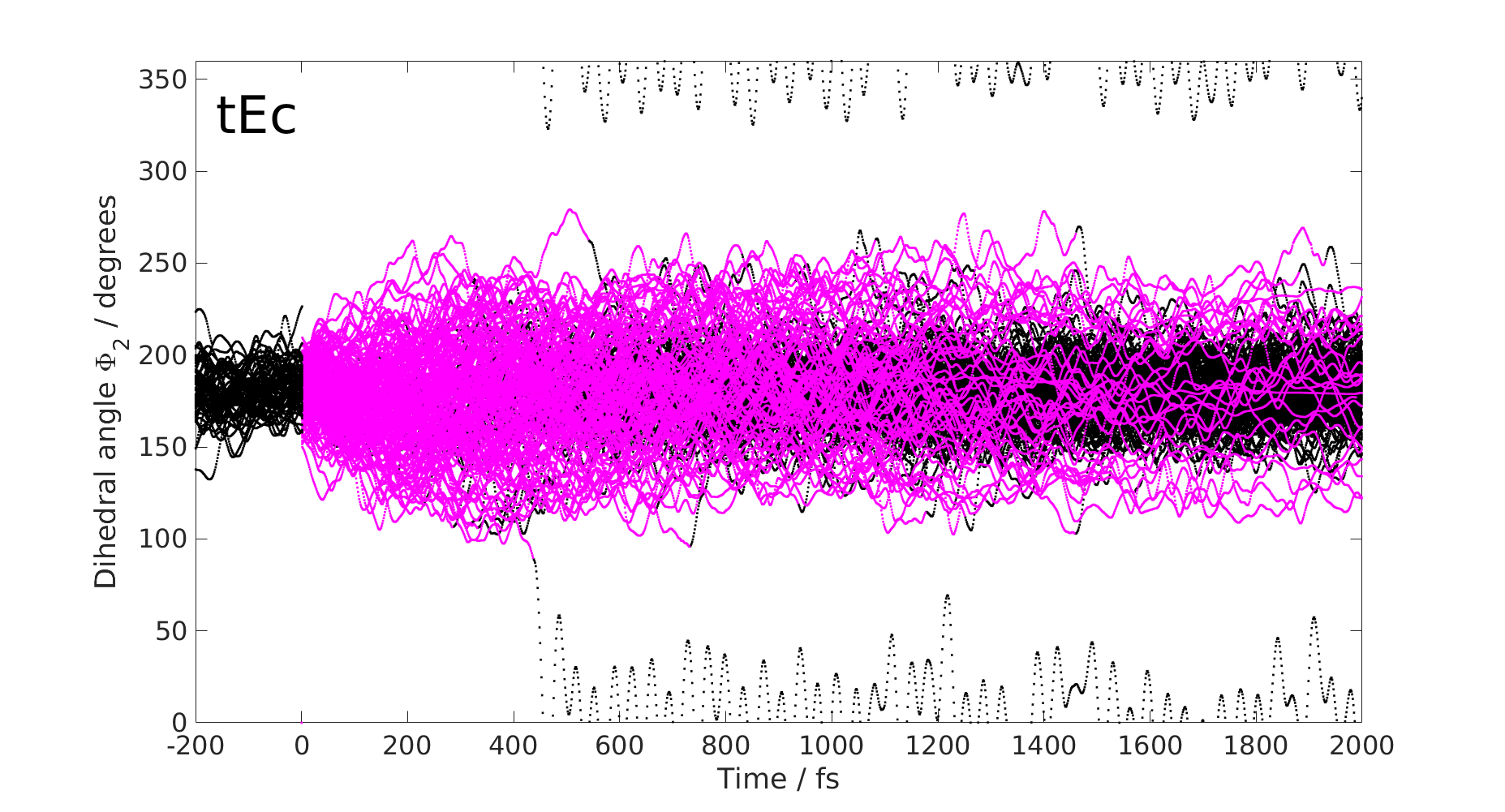}
\includegraphics[scale=0.16]{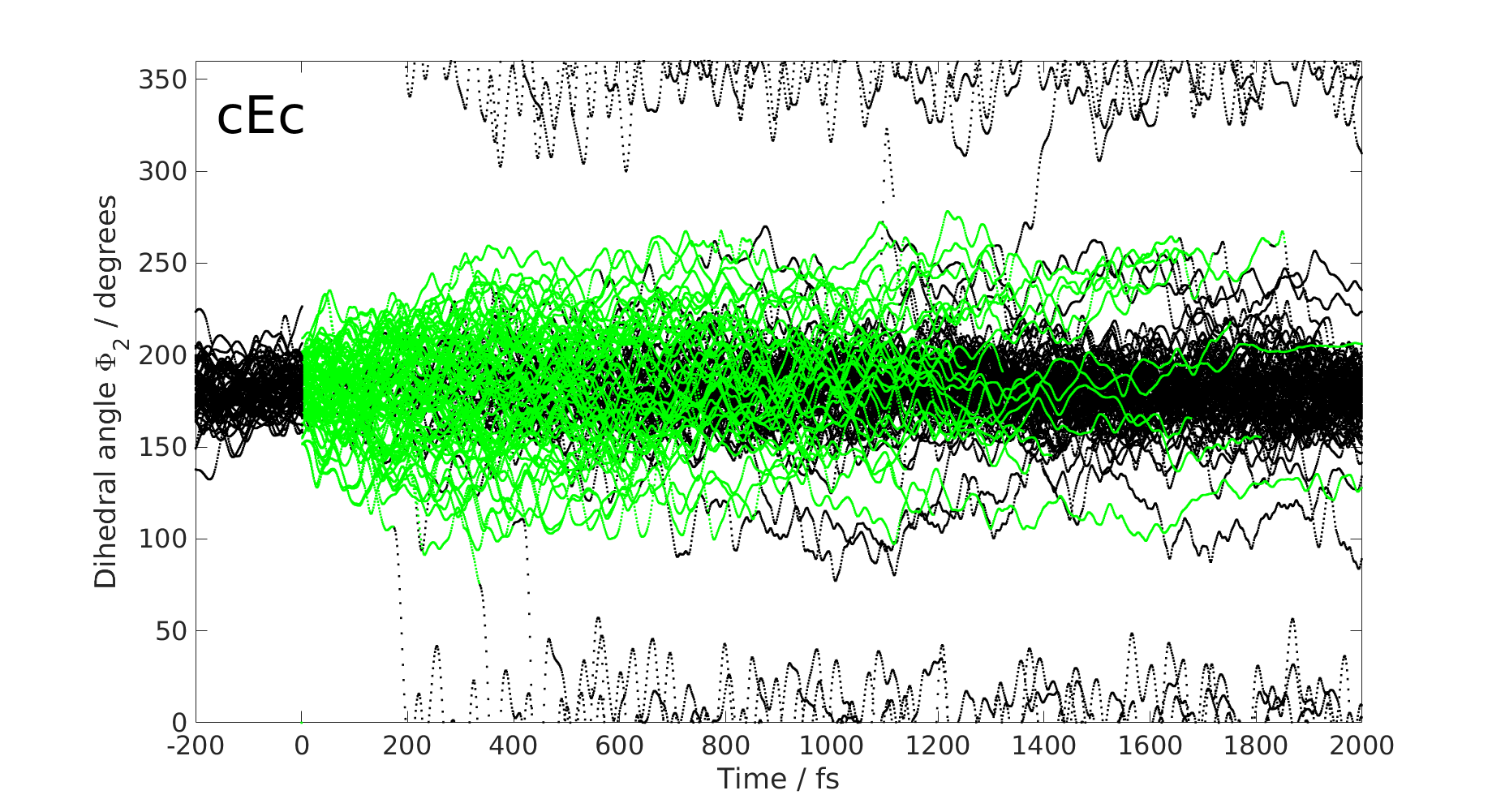}
\caption{\label{dihedrals} Time evolution of the dihedral angle of the central double bond ($\phi_2$) for the different rotamers. Color indicates that the trajectory is currently in the excited state, black indicates the trajectory is in the ground state.} 
\end{figure}

\begin{figure}
\includegraphics[scale=0.37]{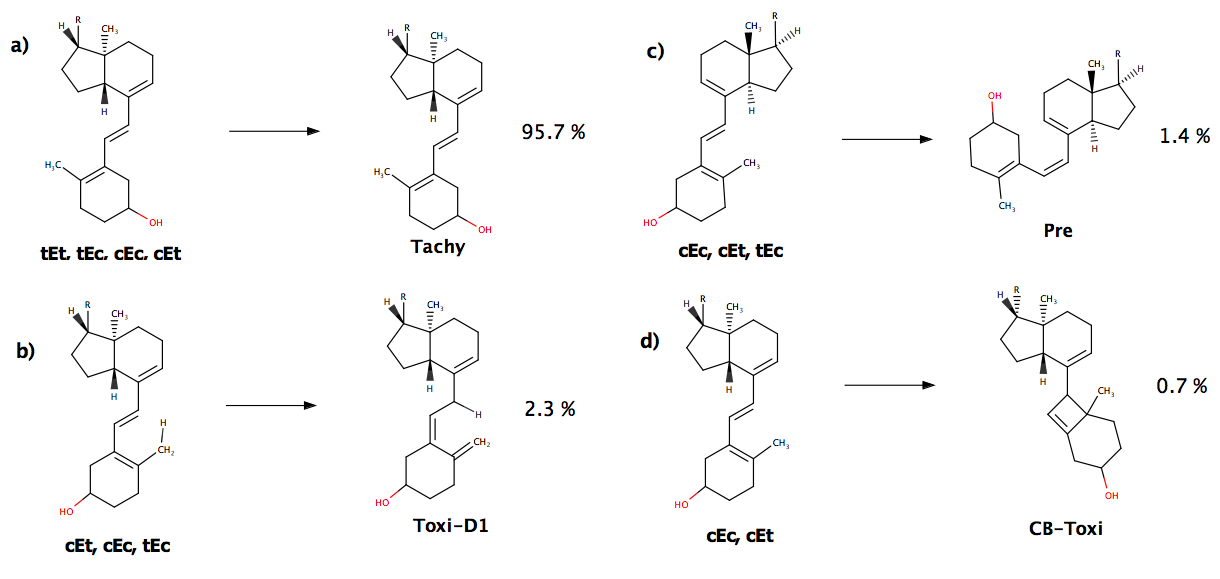}
\caption{\label{overview} Overview of the observed photoreactions of tachysterol: a) unreactive, b) [1,5]-sigmatropic hydrogen transfer (toxisterol D1 formation), c) hula-twist double bond isomerization (previtamin D formation), d) thermal 2+2 electrocyclization (toxisterol-CB formation). Only the main rotamer in which the reaction occurs is shown, other rotamers in which the reaction was observed are listed below the reactant. The overall percentage for each reaction channel is given.}
\end{figure}

To study the excited state dynamics of Tachy, 442 snapshot geometries from REMD were used as starting structures and initiated in the first excited state and propagated by the TDDFT-SH algorithm.
After photoexcitation, we observe a strong twisting of the central double bond in all trajectories (Fig. \ref{dihedrals}). Twisting induces four different relaxation channels with four distinct reaction products, summarized in Fig. \ref{overview} and Table \ref{rotamer_dist_2}.
Most trajectories {(95.7 \%)} decay to the ground state without chemical transformation (unreactive channel), which can be seen by the dihedral angle $\phi_2$ returning to its typical ground state value after relaxation to the ground state (Fig. \ref{dihedrals}). For the unreactive trajectories, the twisting around $\phi_2$ leads to a change in distribution of the molecules in the $\phi_3/\phi_1$-space, which becomes obvious from the comparison of the distribution of structures at time zero (Fig. \ref{histogram}, left) with the distribution at the time of the surface hop (Fig. \ref{histogram}, right). 
{Two picoseconds after excitation, when all trajectories relaxed to the ground state, we note that a large interconversion between rotamers has occurred, resulting in a slightly different distribution of the absolute amounts of the rotamers (tEt: 25.8 \%, cEt: 10.4 \%, tEc:  46.4 \%, cEc: 13.1 \%) than at in the beginning of the TDDFT-SH simulation (Table \ref{rotamer_dist_2}). While the absolute amounts of tEt and cEt are almost the same as in the beginning of the simulation, the amount of tEc increased from 177 structures to 205, whereas the amount of cEc structures decreased from 100 to 58 structures. As we see below, one reason for the strong decrease in cEc structures is its higher photoreactivity forming Pre and toxisterols, but this only explains about a fifth of the decrease. The remaining difference is due to conversion of cEc conformers to the more stable tEc and tEt conformers as a result of photo relaxation (Table \ref{rotamer_dist_2}, Fig. 5 of the ESI). }

\begin{figure}
\includegraphics[scale=0.4]{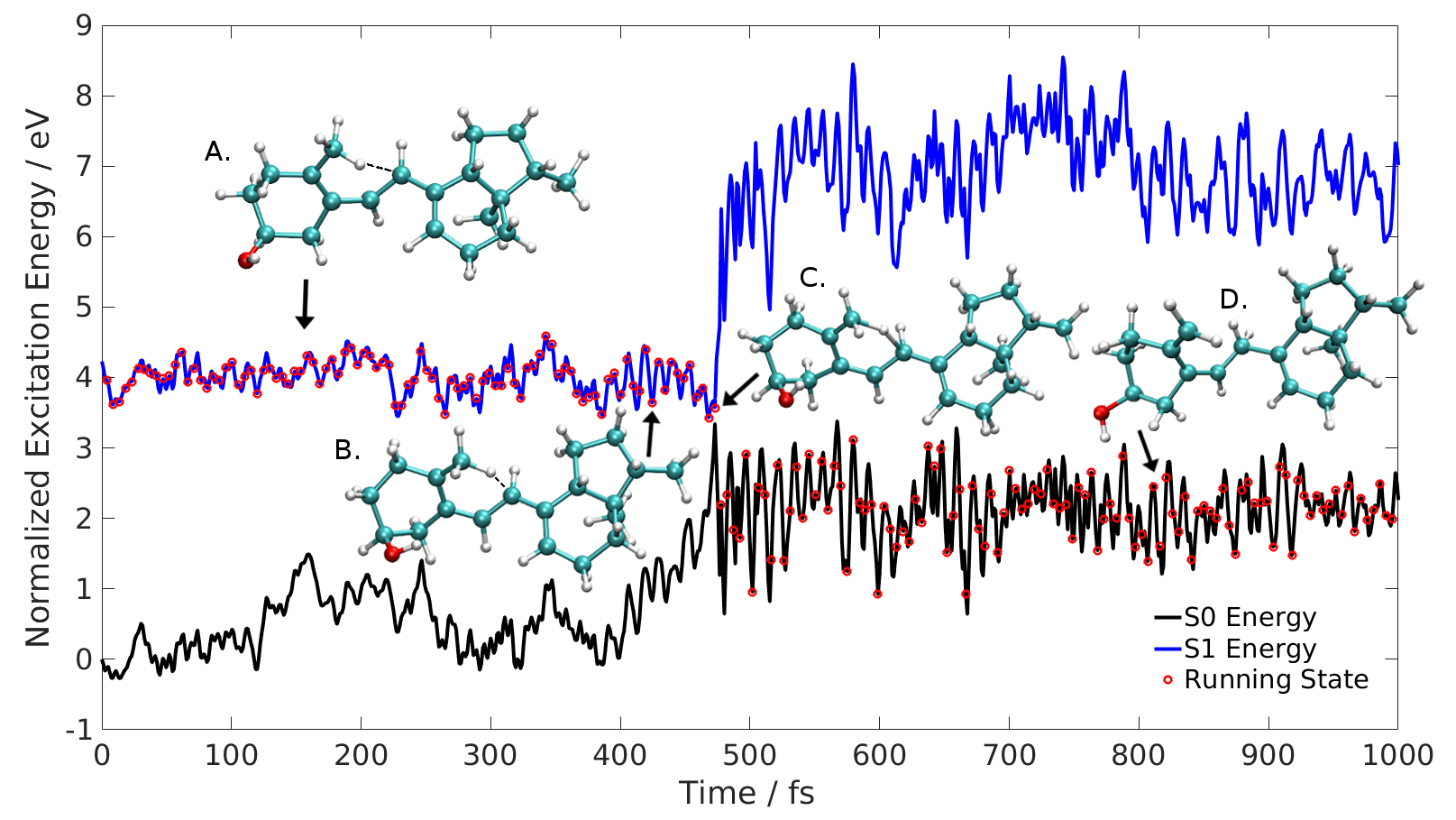}
\caption{\label{htrans} Snapshots along the [1,5]-sigmatropic hydrogen transfer (Toxi-D1 formation), trajectory starts with cEc rotamer as defined in Fig. \ref{rotamers}.}
\end{figure}

\begin{figure}
\includegraphics[scale=0.5]{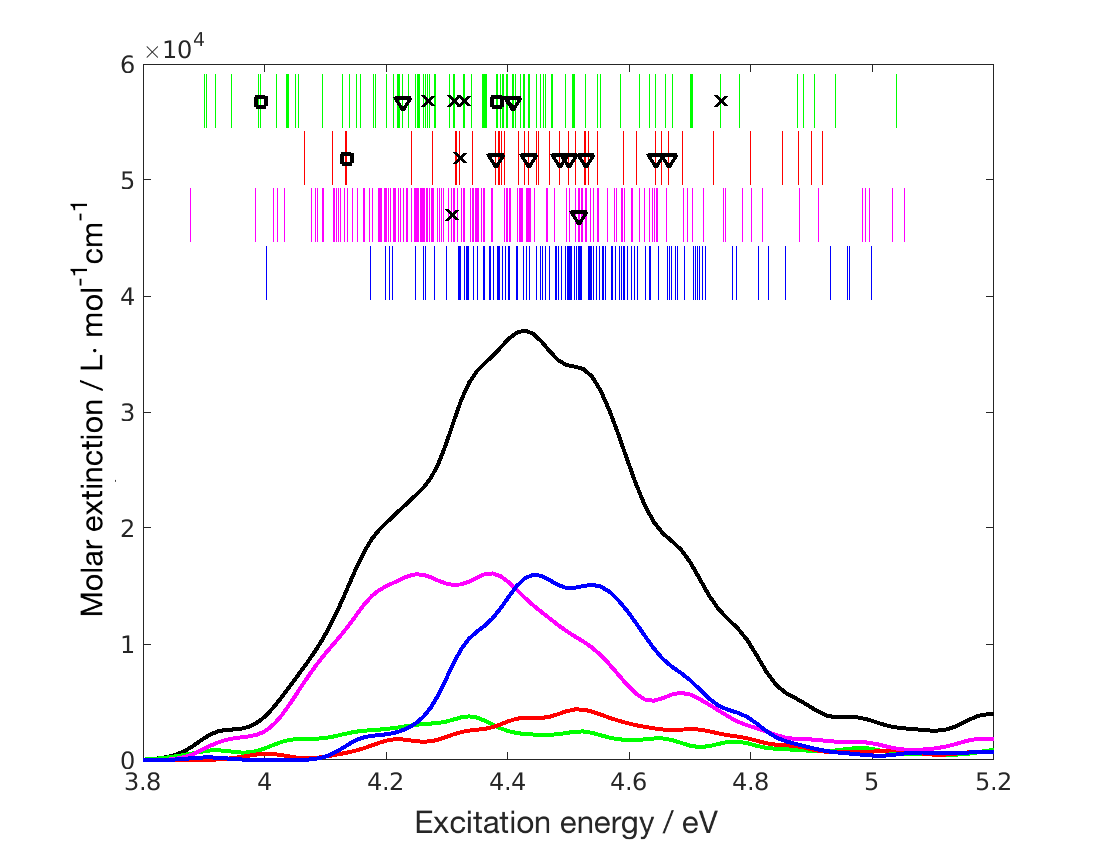}
\caption{\label{excistart} {Excitation energies of the starting structures for the TDDFT surface hopping dynamics are indicated by the bars on top of the spectrum. For comparison, the overall Tachy absorption spectrum (black) and the spectra of the individual rotamers are shown below. The TDPBE0 spectra and excitation energies have been shifted upwards by 0.35 eV to match the absorption peak maximum with the maximum of the experimental spectrum. In both, bars and the spectrum, the color indicates the specific rotamer: tEt: blue, tEc: magenta, cEt: red, cEc: green. Symbols indicate the photoproduct for the specific trajectory. Cross: trans-cis isomerization (product Pre), triangle: hydrogen transfer (Toxisterol D1); square: thermal cyclobutene (CB) formation (product: CB-Toxisterol). No symbol means unreactive trajectory.} }
\end{figure}

As second dominant reaction channel, we observe in {10 trajectories (2.3 \%)} [1,5]-sigmatropic hydrogen transfer from carbon C-19 to carbon C-7, forming the experimentally confirmed\cite{Boomsma1975,Boomsma1977overirradiation} partly deconjugated 
9,10-seco-triene toxisterol D1 (Toxi-D1) (Fig. \ref{overview}b and Fig. \ref{htrans}). Most of the starting structures of these trajectories stem from the cEc and the cEt pool. At the time of the hydrogen transfer, which coincides with time of the surface hop to the ground state, all these trajectories can be found in the cEc or the cEt region (Fig. \ref{histogram}, right). This is expected, since [1,5]-sigmatropic hydrogen transfer requires carbons C-19 and C-7 to be close enough to react, which is only fulfilled by cEc and cEt conformers.
{Relating the hydrogen-transfer trajectories with the initial excitation energy (Fig. \ref{excistart}), we see that this reaction occurs in the center and higher energy region of Tachy's absorption spectrum and is less likely to occur at longer wavelengths.}
   
\begin{figure}
\includegraphics[scale=0.4]{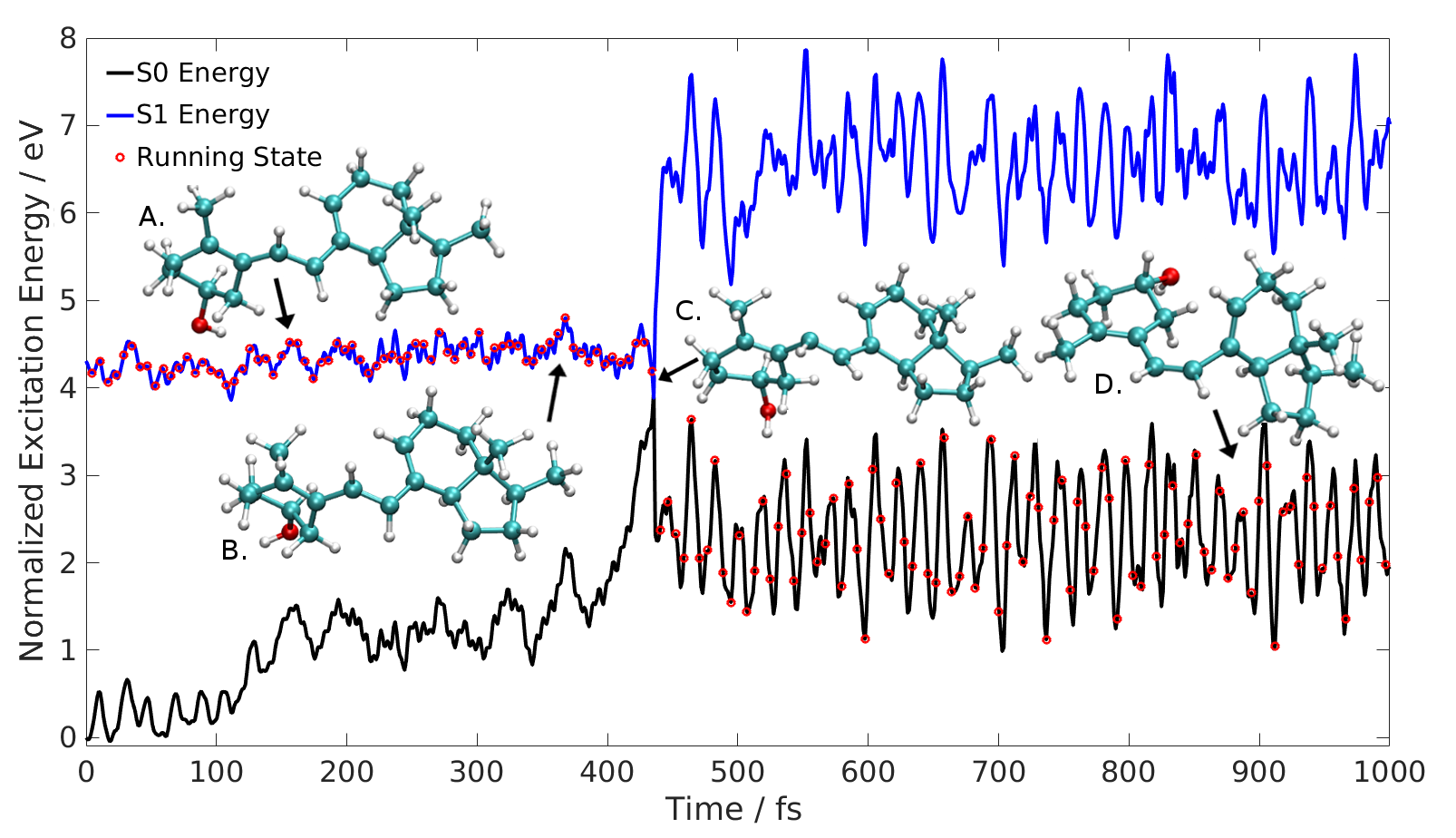}
\caption{\label{hula} Snapshots along the previtamin D formation, via the hula twist, trajectory starts with cEc rotamer as defined in Fig. \ref{rotamers}.}
\end{figure}

\begin{figure}
\includegraphics[scale=0.4]{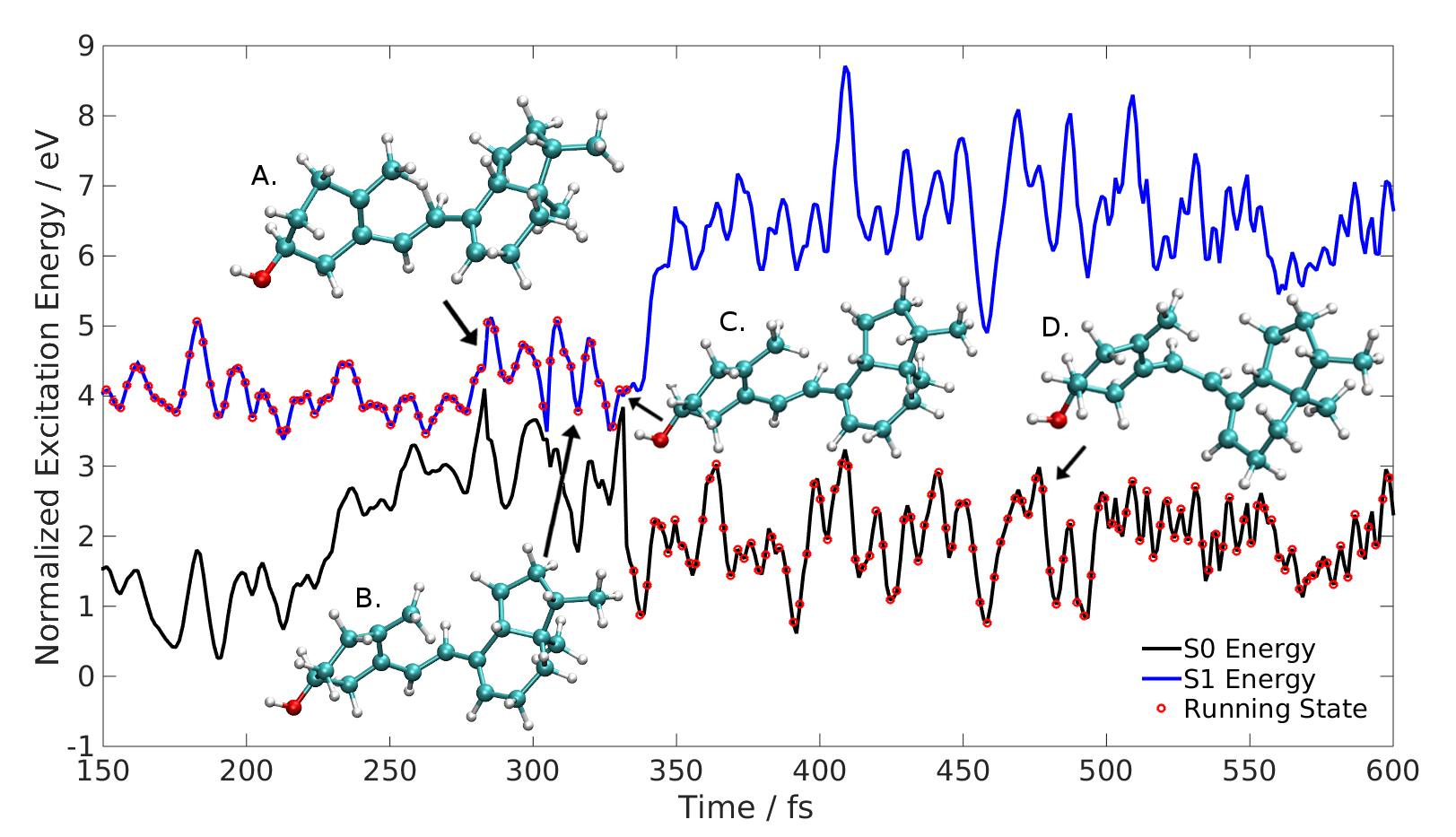}
\caption{\label{htrans_twist} Snapshots along the previtamin D formation, via reversible hydrogen transfer and hula twist, trajectory starts with cEc rotamer as defined in Fig. \ref{rotamers}.}
\end{figure}

\begin{figure}
\includegraphics[scale=0.3]{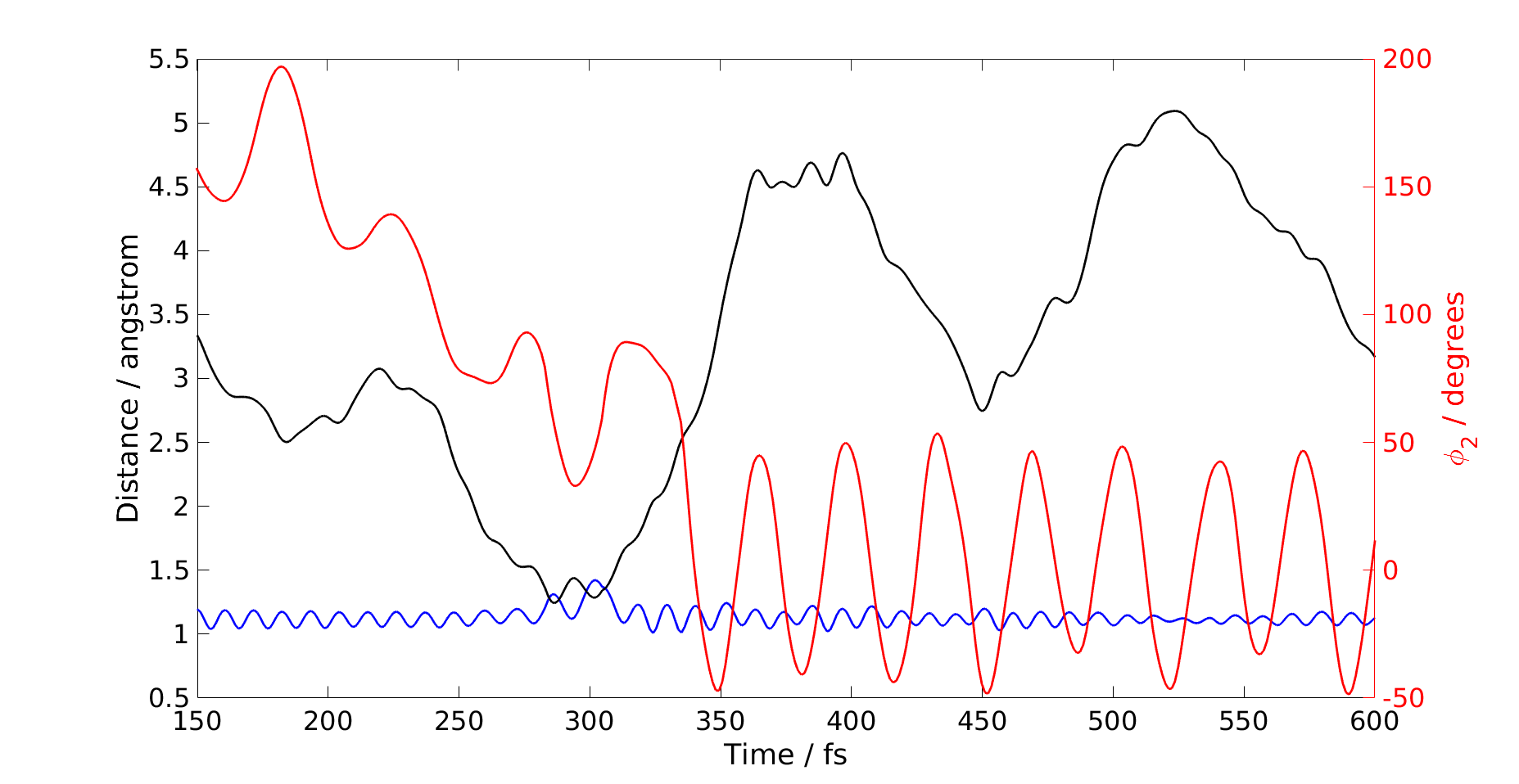}
\caption{\label{dist_dihed} Time evolution of the distance methyl-hydrogen (C19)H and carbon C7 (black), the carbon-hydrogen distance C19 and (C19)H (blue), and H-C6-C7-H dihedral angle (red) of the central double bond of the trajectory shown in Fig. \ref{htrans_twist}. Reversible [1,5]-hydrogen transfer occurs around 300 fs shortly before the double bond isomerizes.}
\end{figure}

As third dominant reaction channel, Pre formation via hula-twist double bound isomerization occurs in 1.4 \% of the trajectories (Fig. \ref{hula} and \ref{htrans_twist}). At the time of the surface hop, we find 5 of 6 trajectories located in the same region where hydrogen transfer occurs (cEc and cEt) and one trajectory in the tEc region (Fig. \ref{histogram}, right). 
In one trajectory we find a reversible [1,5]-hydrogen shift from C-19 to C-7 before the double bond isomerizes (Fig. \ref{htrans_twist} and \ref{dist_dihed}). 
It is not clear if the reversible hydrogen transfer influences the mechanism of the double bond isomerization or if there is simply a competition between these two reaction channels in the cEc and cEt region. 
Further mechanistic investigations are necessary to answer this question.
However, there seems to be a strong preference for trans-cis isomerization in the cEc region, which  
 is consistent with the hypothesis of Saltiel, stating that cEc conformers have higher Pre formation probability \cite{Saltiel2003}.
Together with the fact that cEc absorption band is found at the long wavelength side of the spectrum, this could explain the experimental finding \cite{maessen1983photochemistry,Saltiel2003,andreo2015generation} of increased Pre production when Tachy is excited at wavelengths longer than 300 nm.
{Examining the excitation energies of the starting structures of the previtamin D forming trajectories (Fig. \ref{excistart}), we see that all, except one, trajectories exhibit excitation energies between 4.26--4.32 eV (291 -- 287 nm) close to the cEc absorption peak maximum on the red side of the peak of the overall Tachy absorption spectrum (4.43 eV, 280 nm). Although we do not find a previtamin D forming trajectory with excitation wavelength longer than 300 nm, our simulations exhibit a tendency to form previtamin D at the red side of the Tachy absorption spectrum.}
{Interestingly we also find one Pre forming trajectory at the high energy region at 261 nm (Fig. \ref{excistart}), which is consistent with the observation of Havinga et al. \cite{Havinga1960}, who found Pre formation upon excitation at 253.7 nm. }

\begin{figure}
\includegraphics[scale=0.4]{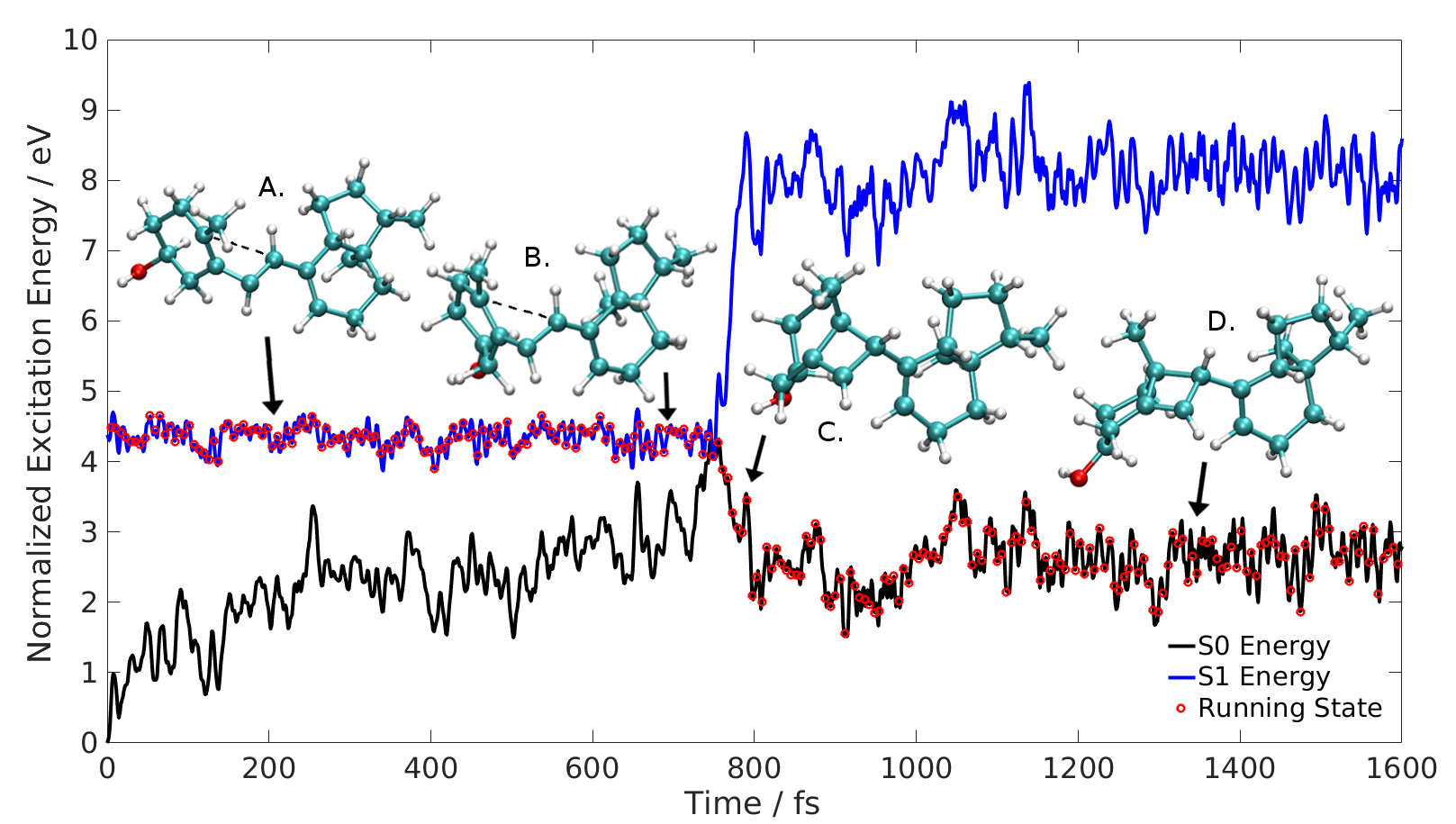}
\caption{\label{cyclobutene} Snapshots along the cyclobutene formation in the hot ground state, trajectory starts with cEt rotamer as defined in Fig. \ref{rotamers}.}
\end{figure}

Lastly, in three trajectories (0.7 \%) we find formation of a cyclobutene toxisterol (CB-Toxi) during a hot ground state reaction (Fig. \ref{cyclobutene}). The four ring is formed by carbons C-5, C-6, C-7, C-10, and shares carbons C-5 and C-10 with the six ring of the steroid A unit. 
To our knowledge, this cyclobutene derivative has only been been found upon direct irradiation of vitamin D\cite{Jacobs1981}. In addition, a different cyclobutene has been characterized by NMR and found upon prolonged irradiation of provitamin D\cite{Boomsma1977overirradiation}.
In this cyclobutene derivative the four ring is formed on the steroid C ring (involving carbons C-6, C-7, C-8, C-9). 
{All three CB-Toxi forming trajectories exhibit excitation energies on the red side of Tachy absorption spectrum, two of them are found at the red tail of the spectrum (Fig. \ref{excistart}). Under the assumption that CB-Toxi can thermally react to previtamin D\cite{Boomsma1977overirradiation}, this could also explain the formation of Pre at long wavelengths.}

Overall, the largest pool of tEc and tEt rotamers behave photochemically inert. With the exception of one trajectory from the tEc region that isomerizes to Pre, we only observe unreactive excited state decay. cEt and cEc rotamers behave photochemically more reactive, both isomers exhibit formation of Pre, Toxi-D1, and CB-Toxi (Fig. \ref{histogram}, right).

Comparing the product distribution with quantum yields of the {Tachy $\rightarrow$ Pre reaction} found in the literature, we find several differences. The Pre quantum yield for excitation at high energies (253.7 nm) in ether was reported to be 7.7 \% \cite{Havinga1960}. 
{In addition it has been found that the trans-cis isomerization only occurs at room temperature and not at 80 K\cite{Havinga1960}. This indicates that thermal energy could be necessary for this process to happen. This energy could be provided by excitation at high energies and could thus explain the Tachy formation upon excitation at 253.7 nm, and indeed we do find one Pre forming trajectory on the blue side of the Tachy absorption spectrum (Fig. \ref{excistart}). However, more research is necessary to answer if Tachy formation is due to excess energy or because of the selective excitation of a specific rotamer.}

The later study of Saltiel\cite{Saltiel2003} achieved 45 \% quantum yield for excitation of Tachy at 313 nm, on the red side of Tachy absorption spectrum. Our simulation only shows 1.4 \% of all trajectories to form Pre, but in our simulations starting structures were randomly chosen from the Boltzmann ensemble and no selection with regard to the excitation energy was made. 
{Furthermore, the study of Havinga et al.\cite{Havinga1960} reports 3.3 \% ring-closure forming Lumi, which did not occur in any of our simulations. However, it seems very unlikely to observe direct Lumi formation from Tachy in a one-photon process. Due to the trans-conformation of the central double bond in Tachy, carbon atoms C-9 and C-10 are too far apart to form the bond necessary for ring-closure. This could only occur at very strong twisting of the central double bond, which would most likely trigger direct relaxation to the ground state. More likely, the observed Lumi formation stems from a two-step process forming first Pre and then, after a second photon is absorbed by Pre, ring-closure from Pre to Lumi (Tachy $\rightarrow$ Pre $\rightarrow$ Lumi).}  
Another difference between experiments and our simulation is the formation of large amounts of Toxi-D1 and CB-Toxi, which are not mentioned as products in the work of Saltiel \cite{Saltiel2003}.
Possible reasons for the differences of our gas phase simulations with experiments in solution might be that a) in our simulation no specific wavelength was assumed for the initial excitation since the starting structures were taken equally from the ensemble of structures from the 300 K Boltzmann distribution, and b) the neglect of the solvent in our simulations might also have an impact on the product distribution. {This could be due to either a different conformer distribution in solution or due to trapping of intermediates by the solvents.} 
{c) It cannot be excluded that processes involving absorption of two or more photons occurred in the cited experimental studies. Since we restrict our simulations to one-photon processes, by definition these reaction pathways cannot be described.
d) Furthermore, the limited number or reactive trajectories (19 out of 442) possibly introduces a sampling error in the product distribution. Using 442 trajectories, with a 90 \% confidence, the maximum margin error ($\epsilon$) in the product distribution amounts to only a few percent (1.3--5.2 \%) for most products, with maximum errors of 6.6 and 7.5 \% for the rotamers cEt and tEc, respectively (Table~\ref{rotamer_dist_2}). This small margin error indicates that a semi-quantitative prediction of the branching ratios is possible with 442 trajectories, but possibly other photoproducts with small quantum yield might not have been detected by our simulations.}

To assess the accuracy of TDPBE0/SVP potential energy surfaces along the excited state decay, we computed CC2 potential energy surfaces of each of the three reactive decay channels (Fig. 2-4, ESI). In all cases, CC2 T2-amplitudes of S$_1$ are 6 \% on average and always below 10 \%. The differences in the predicted $\omega_1$ by TDPBE0/SVP and CC2/TZVP are 0.24 eV on average, during the excited state dynamics. CC2 and TDPBE0 both predict S$_2$-S$_1$ energy gaps to be 1 eV on average, which indicates small involvement of the second excited state in the dynamics.

\begin{figure}
\includegraphics[scale=0.8]{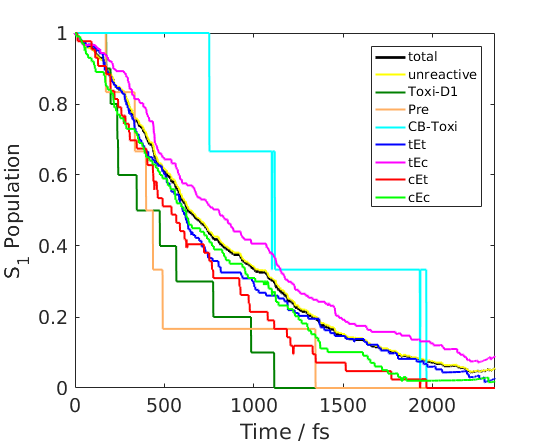}
\caption{\label{decay} Population of the first excited state as a function of time. A mono-exponential fit of the total decay gives a time constant of 882 fs. }
\end{figure}

Analyzing the the evolution of the S$_1$ population during excited state dynamics (Fig. \ref{decay}), we find a mono-exponential decay of S$_1$ with a time constant of $\approx$ 880 fs. This is longer than the lifetimes predicted by TDDFT-SH of Pro (265 fs) and Pre (534 fs) \cite{Tapavicza2011}. It appears that the mainly unreactive tEc rotamers exhibit the longest excited state lifetimes, whereas the more reactive cEt rotamers exhibit the shortest lifetimes. 
Furthermore, it can be seen that the reactive trajectories forming Pre and Toxi-D1 exhibit the shortest lifetimes.
Compared to its analogue model compound trans-hexatriene, which exhibits a lifetime of 190$\pm$30 fs \cite{Garavelli1997_trans-hexatriene,Anderson2000_cis_trans_hexatriene}, controlled by intramolecular vibrational energy redistribution, Tachy has much longer excited state lifetimes.
This could be due to the decreased flexibility of the conjugated double bond system due the constraints of the steroid rings A and C (Fig. \ref{rotamers}). To answer this question additional studies are necessary.

\section{The role of tachysterol}
Tachysterol is formed by cis-trans isomerization of Pre under conditions where UV radiation of high energy is available\cite{Dmitrenko1999,Tapavicza2011}. This is the case at small zenith angles, for instance at noon or in summer. At these conditions also the Pro ring-opening forming Pre is dominant.
Tachy covers the absorption spectra of Pro and Lumi, which both can lead to Pre at wavelengths shorter than 320 nm. 
If large quantities of Tachy accumulate, the increased absorption of UV light by Tachy reduces the probability of Pre formation by Pro and Lumi ring-opening simply by a competition over the available photons.
At the same time, Tachy conformers that absorb high energy UV radiation are mainly the photochemically inert tEt and tEc rotamers, which dissipate radiation energy mainly through unsuccessful double-bond twisting in an unreactive excited state decay. This behaviour indicates a sun screen effect of tachysterol at higher excitation energies, throttling Pre formation and eventually vitamin D production.
In addition, the minor formation of toxisterols also leads to a reduction of Pre formation.
In contrast, under increased solar zenith angles, where the global spectral flux is shifted to longer wavelengths, the contribution of cEc rotamers to the total absorption is increased. Together with the increased tendency of cEc to form Pre, this constitutes a source of Pre in winter and in the early morning as well as in the evening, which has recently been confirmed experimentally\cite{andreo2015generation}.
This is against the common belief that Pre can only be formed at wavelengths shorter than 320 nm. However, this statement was based on the assumption that Pre originates from Pro ring-opening, which has its maximum efficiency at wavelengths of approximately 290 nm.

The largest contribution to toxisterol formation is found by cEt and cEc rotamers. Compared to previous simulations of Pre\cite{Tapavicza2011}, Tachy exhibits much larger Toxi formation than Pre. Thus, it appears that Tachy is an important intermediate in the degradation of DPI to Toxisterols.
However, the CB-Toxi found in our study is expected to form previtamin D through thermal ring-opening \cite{Boomsma1977overirradiation}  and possibly constitutes another Pre reservoir, which can be tapped thermally.
{Together with the finding that CB-Toxi is formed at the red tail of Tachy's absorption spectrum, this is another explanation of Pre formation upon excitation with long wavelengths.}
In addition, it has been found that excitation at very short wavelengths (253.7 nm) can also lead Pre formation from Tachy \cite{Havinga1960}. Disregarding the question whether this is due to excess energy or not, this pathway is not important for natural Vita photosynthesis since the amounts of light at this wavelength is negligible at sea level\cite{green1974middle}.

\section{Conclusion}
Our study shows that the conformational dependency of Tachy absorption and the different photoreactivity of the rotamers are necessary to explain the experimentally found wavelength dependent photochemistry.
The applied theoretical methods are able to give a consistent and conclusive explanation of these phenomena.
Highly accurate excitation energies can be obtained from CC2 employing the TZVP basis set. However, TDPBE0 potential energy surfaces are consistent with CC2 and allow direct on-the-fly simulations of a large number of trajectories.
Our calculations give information about the contributions of each rotamer to the broad absorption band and strongly support the hypothesis that cEc is more likely to form Pre. Our findings explain why this reaction is enhanced at the red side of the absorption spectrum \cite{Saltiel2003,andreo2015generation}. 

The rotamer resolved spectrum can possibly be used to refine photo kinetic modelling studies and give more insights about vitamin D self-regulation. 
Our study also shows that photoexcitation distorts the equilibrium of rotamers. 
This indicates that the distribution of rotamers is a function of the irradiation conditions and consequently needs to be taken into account in photo kinetic modelling of the system. 
{In particular, this could be important at high radiation intensity, where the distorted equilibrium of rotamers is re-excited by incoming photons.} This could also contribute to the difference of the action spectra for unexposed and previously irradiated skin samples \cite{vanDijk2016}.
{However, the timescale of rotational isomerization is expected to be in the order of 100 picoseconds\cite{Anderson1999} and cannot be assessed by our simulations with a total of 2 ps simulation time. Furthermore, the description of this process requires the inclusion of a chemical environment that mediates energy dissipation. Further simulations on longer timescales are necessary to assess the effect of the distorted rotamer distribution quantitatively.}
Another question is how the cellular membrane influences the distribution of rotamers and the photo reactivity. 
A different double bond isomerization probability can be expected if a solvent or the cell membrane is able to change the population of the cEc basin in the ground state.
In addition the presence of the solvent might also influence the hydrogen transfer reaction and might be responsible for the quantitative differences of between the experimentally determined quantum yield and the predicted ones.
{To get a more accurate description of the hydrogen shift reactions, it would be interesting to apply more sophisticated models that include quantum effects\cite{Tuckerman1997,Zimmermann2010}.}

Our results also opens new mechanistic questions about photochemical reactions of tachysterol.
We observed a strong interplay or competition between hydrogen transfer and trans-cis isomerization.
This could indicate that the C19 methyl group constitutes an important functional group in the photochemical control of Tachy. Future studies are necessary to investigate this question. A possibility to assess this question could be to investigate tachysterol derivatives with different functional groups.

On the photobiological side, there is still the question about the fate and the role of formed toxisterols.
Do these compounds influence the regulation of the photoequilibrium, will they reenter the photoequilibrium or will they simply degrade to unreactive metabolites? Future theoretical and experimental studies focusing on these compounds could answer these questions.

In summary, due to its large extinction coefficient and mostly unreactive behavior, tachysterol acts mainly as a sun shield suppressing previtamin D formation from Pro and Lumi since it absorbs in the same spectral region. Tachysterol shows stronger toxisterol formation than previtamin D and can thus be seen as the major degradation route of vitamin D. Despite being mostly unreactive, to a small amount tachysterol reacts back to previtamin D, which is favored by cEc rotamers at the red side of the Tachy absorption peak maximum and could therefore constitute a previtamin D reservoir for times where only low energetic UV radiation is present.

To validate the findings of our computational study, it would be interesting to complete our study with experimental time-resolved spectroscopic measurements.  This could also answer the different time-constants between Tachy and trans-hexatriene and answer the questions about the influence of the steroid rings A and C and other functional groups, such as the the C19-methyl group.

\begin{acknowledgement}

We acknowledge financial support from CSULB, the CSUPERB New Investigator (NI) grant, and NIH-BUILD computer grant.

\end{acknowledgement}
\begin{table}[h]
{
\centering
\begin{tabular}{lccccc}
                                                 & {\bf tEt}           & {\bf cEt}              & {\bf tEc}          & {\bf cEc}          & Total          \\ \hline\hline
\multicolumn{1}{l}{REMD}                    & 23192 (26.4\%)& 11479 (13.1\%)     & 39850 (45.3\%)& 13386 (15.2\%)    & 87907 (100 \%) \\ 
\multicolumn{1}{l}{MM\cite{Dmitrenko1997}}         &  13 \%&  6 \%     & 63 \%& 18 \%    & 100 \% \\ 
\multicolumn{1}{l}{Spectrum}     & 423 (31.4 \%) & 170 (12.6 \%)         & 556 (41.2 \%)   & 200 (14.8 \%)      & 1349 (100 \%)   \\ 
\multicolumn{1}{l}{TDDFT-SH }     &  123 (27.8 \%) & 42 (9.7 \%)         & 177 (40.0 \%)   & 100 (22.6 \%)      & 442 (100 \%)  \\ 
\multicolumn{1}{l}{Spec. Overl.}     & 4.9 \% & 4.8 \%         & 57.5 \%   & 32.8 \%      & 100 \%  \\ \hline \hline
\multicolumn{5}{l}{Photoproducts}   & percentage $\pm$ $\epsilon$                     \\ \hline
\multicolumn{1}{c}{Tachy}                      & 123 (27.8 \%) & 33 (7.5 \%)        & 175 (39.6 \%) & 92 (20.8 \%)      &  423 (95.7 \% $\pm$ 3.1\%) \\ 
\multicolumn{1}{c}{tEt}                      & 55 (12.4 \%) & 10 (2.3 \%)        & 27 (6.1 \%) & 22 (5.0 \%)      & 114 (25.8 \% $\pm$6.6 \%) \\ 
\multicolumn{1}{c}{cEt}                      & 19 (4.3 \%) & 7 (1.6 \%)        & 9 (2.0 \%) & 11 (2.5 \%)      & 46 (10.4 \% $\pm$4.6 \%) \\ 
\multicolumn{1}{c}{tEc}                      & 32 (7.2 \%) & 15 (3.4 \%)        & 121 (27.4 \%) & 37 (8.4 \%)      & 205 (46.4 \% $\pm$7.5 \%) \\ 
\multicolumn{1}{c}{cEc}                      & 17 (3.8 \%) & 1 (0.2 \%)        & 18 (4.1 \%) & 22 (5.0 \%)      & 58 (13.1 \% $\pm$5.1 \%) \\ 
\multicolumn{1}{c}{Toxi-D1}                    &     --        & 7 (1.6 \%)         & 1 (0.2 \%)    & 2 (0.5)           & 10 (2.3 \% $\pm$2.3 \%)    \\ 
\multicolumn{1}{c}{Pre}                        &     --        & 1 (0.2 \%)         & 1 (0.2 \%)    & 4 (0.9 \%)        &  6 (1.4 \% $\pm$1.8 \%)    \\ 
\multicolumn{1}{c}{CB-Toxi}                    &     --        & 1 (0.2 \%)         &   --          & 2 (0.5 \%)        &  3 (0.7 \% $\pm$1.3 \%)    \\ \hline \hline
\end{tabular}
\caption{\label{rotamer_dist_2}Rotamer distribution at 300 K obtained from REMD and molecular mechanics at 298 K (MM) \cite{Dmitrenko1997}, distribution of TDDFT-SH starting structures (TDDFT-SH), and distribution of photoproducts according to the conformation of their starting structure. The number of structures or trajectories are given with their percentages in parenthesis; in case of MM and the normalized spectral overlap with the standardized spectral irradiance \cite{astm} (Spec. Overl.) percentages are given. {For the total percentages of the photoproducts the statistical maximum margin error ($\epsilon$) due to the finite number of trajectories with 90 \% confidence is given, according to the weak law of large numbers\cite{Khintchine1929,grinstead2012introduction}. The rotamer distribution has been determined 2 ps after initial excitation.}}}
\end{table}
\begin{table}[h]
\centering
\begin{tabular}{lllllll}
   & \multicolumn{2}{c}{TDPBE0} & \multicolumn{4}{c}{CC2} \\\hline
   & \multicolumn{1}{c}{SVP} & \multicolumn{1}{c}{TZVP}&\multicolumn{2}{c}{SVP} & \multicolumn{2}{c}{TZVP}\\\
   &   $\omega$\,\,\,\,\,\,\,  ($f$)           &  $\omega$\,\,\,\,\,\,\,  ($f$)             & $\omega$\,\,\,\,\,\,\,  ($f$) & $t2$ \%    & $\omega$\,\,\,\,\,\,\,  ($f$)  & $t2$ \%  \\\hline
cEc&   3.98 (0.686)                  &   3.86    (0.635)                & 4.59  (0.786)                 &5.56 &  4.39 (0.711)                 &5.82\\
   &   4.98 (4.20 $\times 10^{-3}$)  &   4.88    (4.31$\times 10^{-3}$) & 5.62  (4.06$\times 10^{-3}$)  &7.87 &  5.42 (4.20$\times$10$^{-3}$) &8.16\\\hline
cEt&   4.20 (0.732)                  &   4.12    (0.707)                & 4.83  (0.855)                 &5.77 &  4.68 (0.817)                 &6.07\\
   &   5.16 (3.95$\times 10^{-2}$)   &   5.09    (3.72$\times 10^{-2}$) & 5.82  (3.81$\times 10^{-2}$)  &8.20 &  5.64 (3.43$\times$10$^{-2}$) &8.47\\\hline
tEc&   4.09 (0.789)                  &   3.99    (0.750)                & 4.70  (0.910)                 &5.68 &  4.53 (0.847)                 &5.93\\
   &   5.12 (3.36$\times 10^{-2}$)   &   5.04    (2.58$\times 10^{-2}$) & 5.76  (3.49$\times 10^{-2}$)  &8.22 &  5.58 (2.28$\times$10$^{-2}$) &8.56\\\hline
tEt&   4.26 (1.07)                   &   4.19    (1.04)                 & 4.87  (1.25)                  &5.94 &  4.73 (1.21)                  &6.21 \\
   &   5.25 (3.64$\times 10^{-4}$)   &   5.19    (1.56$\times 10^{-3}$) & 5.85  (5.52$\times 10^{-4}$)  &8.88 &  5.68 (6.95$\times$10$^{-4}$) &9.12 \\ \hline
\caption{\label{opt_exci} Calculated lowest two excitation energies $\omega$ (eV) for the ground state structures optimized by RI-PBE/SVP. Oscillator strengths in atomic units (length gauge) are given in parenthesis. For CC2, the percentage of the coupled cluster $t2$-amplitudes are given.}
\end{tabular}
\end{table}
\begin{table}[h]
\centering
\begin{tabular}{lccccccccc}
&\multicolumn{3}{c}{$\lambda_{\text{max}}$ / eV }&\multicolumn{3}{c}{$\epsilon_{\text{max}}$  / 10$^4$Lmol$^{-1}$cm$^{-1}$}& \multicolumn{3}{c}{FWHM / eV}\\\hline
&TDDFT&CC2&exp.$^a$&TDDFT&CC2&exp.$^a$&TDDFT&CC2&exp.$^b$\\\hline
tEt&4.092&4.714&-&1.597&1.716&-&0.344&0.372&--\\
cEt&4.161&4.824&-&0.436&0.412&-&0.451&0.539&--\\
tEc&4.025&4.558&-&1.608&1.645&-&0.483&0.542&--\\
cEc&3.987&4.592&-&0.376&0.346&-&0.460&0.694&--\\
Total&4.078&4.679&4.428&3.702&3.875&2.8&0.376&0.547&0.692\\\hline
\end{tabular}
\caption{\label{spec_compare} Comparison between spectra calculated by TDDFT and CC2. $^a$Values from Dmitrenko et al. \cite{Dmitrenko1999}, measured in ethanol. $^b$ Estimated from MacLaughlin et al.\cite{MacLaughlin1982}.}
\end{table}
\clearpage
\clearpage
\bibliography{papers}

\providecommand*\mcitethebibliography{\thebibliography}
\csname @ifundefined\endcsname{endmcitethebibliography}
  {\let\endmcitethebibliography\endthebibliography}{}
\begin{mcitethebibliography}{96}
\providecommand*\natexlab[1]{#1}
\providecommand*\mciteSetBstSublistMode[1]{}
\providecommand*\mciteSetBstMaxWidthForm[2]{}
\providecommand*\mciteBstWouldAddEndPuncttrue
  {\def\EndOfBibitem{\unskip.}}
\providecommand*\mciteBstWouldAddEndPunctfalse
  {\let\EndOfBibitem\relax}
\providecommand*\mciteSetBstMidEndSepPunct[3]{}
\providecommand*\mciteSetBstSublistLabelBeginEnd[3]{}
\providecommand*\EndOfBibitem{}
\mciteSetBstSublistMode{f}
\mciteSetBstMaxWidthForm{subitem}{(\alph{mcitesubitemcount})}
\mciteSetBstSublistLabelBeginEnd
  {\mcitemaxwidthsubitemform\space}
  {\relax}
  {\relax}

\bibitem[Dixon et~al.(2005)Dixon, Deo, Wong, Slater, Norman, Bishop, Posner,
  Ishizuka, Halliday, Reeve, and Mason]{Dixon2005}
Dixon,~K.; Deo,~S.; Wong,~G.; Slater,~M.; Norman,~A.; Bishop,~J.; Posner,~G.;
  Ishizuka,~S.; Halliday,~G.; Reeve,~V.; Mason,~R. \emph{J. Steroid Biochem.
  Mol. Biol.} \textbf{2005}, \emph{97}, 137 -- 143\relax
\mciteBstWouldAddEndPuncttrue
\mciteSetBstMidEndSepPunct{\mcitedefaultmidpunct}
{\mcitedefaultendpunct}{\mcitedefaultseppunct}\relax
\EndOfBibitem
\bibitem[Wang(2009)]{Wang2009}
Wang,~S. \emph{Nutr. Res. Rev.} \textbf{2009}, \emph{22}, 188\relax
\mciteBstWouldAddEndPuncttrue
\mciteSetBstMidEndSepPunct{\mcitedefaultmidpunct}
{\mcitedefaultendpunct}{\mcitedefaultseppunct}\relax
\EndOfBibitem
\bibitem[Wang et~al.(2008)Wang, Pencina, Booth, Jacques, Ingelsson, Lanier,
  Benjamin, D'Agostino, Wolf, and Vasan]{Wang29012008}
Wang,~T.~J.; Pencina,~M.~J.; Booth,~S.~L.; Jacques,~P.~F.; Ingelsson,~E.;
  Lanier,~K.; Benjamin,~E.~J.; D'Agostino,~R.~B.; Wolf,~M.; Vasan,~R.~S.
  \emph{Circulation} \textbf{2008}, \emph{117}, 503--511\relax
\mciteBstWouldAddEndPuncttrue
\mciteSetBstMidEndSepPunct{\mcitedefaultmidpunct}
{\mcitedefaultendpunct}{\mcitedefaultseppunct}\relax
\EndOfBibitem
\bibitem[Chen et~al.(2000)Chen, Persons, Lu, Mathieu, and Holick]{Chen2000}
Chen,~T.~C.; Persons,~K.~S.; Lu,~Z.; Mathieu,~J.~S.; Holick,~M.~F. \emph{J.
  Nutr. Biochem.} \textbf{2000}, \emph{11}, 267 -- 272\relax
\mciteBstWouldAddEndPuncttrue
\mciteSetBstMidEndSepPunct{\mcitedefaultmidpunct}
{\mcitedefaultendpunct}{\mcitedefaultseppunct}\relax
\EndOfBibitem
\bibitem[Tuckey et~al.(2014)Tuckey, Slominski, Cheng, Chen, Kim, Xiao, and
  Li]{Tuckey2014}
Tuckey,~R.~C.; Slominski,~A.~T.; Cheng,~C.~Y.; Chen,~J.; Kim,~T.-K.; Xiao,~M.;
  Li,~W. \emph{Int. J. Biochem. Cell Biol.} \textbf{2014}, \emph{55}, 24 --
  34\relax
\mciteBstWouldAddEndPuncttrue
\mciteSetBstMidEndSepPunct{\mcitedefaultmidpunct}
{\mcitedefaultendpunct}{\mcitedefaultseppunct}\relax
\EndOfBibitem
\bibitem[Havinga et~al.(1960)Havinga, de~Kock, and Rappoldt]{Havinga1960}
Havinga,~E.; de~Kock,~R.~J.; Rappoldt,~M.~P. \emph{Tetrahedron Lett.}
  \textbf{1960}, \emph{11}, 276\relax
\mciteBstWouldAddEndPuncttrue
\mciteSetBstMidEndSepPunct{\mcitedefaultmidpunct}
{\mcitedefaultendpunct}{\mcitedefaultseppunct}\relax
\EndOfBibitem
\bibitem[MacLaughlin et~al.(1982)MacLaughlin, Anderson, and
  Holick]{MacLaughlin1982}
MacLaughlin,~J.; Anderson,~R.; Holick,~M. \emph{Science} \textbf{1982},
  \emph{216}, 1001--1003\relax
\mciteBstWouldAddEndPuncttrue
\mciteSetBstMidEndSepPunct{\mcitedefaultmidpunct}
{\mcitedefaultendpunct}{\mcitedefaultseppunct}\relax
\EndOfBibitem
\bibitem[Albright et~al.(1939)Albright, Sulkowitch, and
  Bloomberg]{albright1939comparison}
Albright,~F.; Sulkowitch,~H.~W.; Bloomberg,~E. \emph{J. Clin. Invest.}
  \textbf{1939}, \emph{18}, 165\relax
\mciteBstWouldAddEndPuncttrue
\mciteSetBstMidEndSepPunct{\mcitedefaultmidpunct}
{\mcitedefaultendpunct}{\mcitedefaultseppunct}\relax
\EndOfBibitem
\bibitem[Suda et~al.(1970)Suda, Hallick, DeLuca, and Schnoes]{suda197025}
Suda,~T.; Hallick,~R.; DeLuca,~H.~F.; Schnoes,~H.~K. \emph{Biochemistry}
  \textbf{1970}, \emph{9}, 1651--1657\relax
\mciteBstWouldAddEndPuncttrue
\mciteSetBstMidEndSepPunct{\mcitedefaultmidpunct}
{\mcitedefaultendpunct}{\mcitedefaultseppunct}\relax
\EndOfBibitem
\bibitem[Chen et~al.(2000)Chen, Persons, Lu, Mathieu, and Holick]{Chen2000267}
Chen,~T.~C.; Persons,~K.~S.; Lu,~Z.; Mathieu,~J.~S.; Holick,~M.~F. \emph{J.
  Nutr. Biochem} \textbf{2000}, \emph{11}, 267 -- 272\relax
\mciteBstWouldAddEndPuncttrue
\mciteSetBstMidEndSepPunct{\mcitedefaultmidpunct}
{\mcitedefaultendpunct}{\mcitedefaultseppunct}\relax
\EndOfBibitem
\bibitem[Holick(2007)]{Holick2007}
Holick,~M.~F. \emph{N. Engl. J. Med.} \textbf{2007}, \emph{357}, 266--281\relax
\mciteBstWouldAddEndPuncttrue
\mciteSetBstMidEndSepPunct{\mcitedefaultmidpunct}
{\mcitedefaultendpunct}{\mcitedefaultseppunct}\relax
\EndOfBibitem
\bibitem[Chen et~al.(2010)Chen, Lu, and Holick]{Chen2010}
Chen,~T.~C.; Lu,~Z.; Holick,~M.~F. Photobiology of Vitamin D. Nutrition and
  Health: Vitamin D. 2010; p~35\relax
\mciteBstWouldAddEndPuncttrue
\mciteSetBstMidEndSepPunct{\mcitedefaultmidpunct}
{\mcitedefaultendpunct}{\mcitedefaultseppunct}\relax
\EndOfBibitem
\bibitem[Holick et~al.(1981)Holick, MacLaughlin, and Doppelt]{Holick1981}
Holick,~M.~F.; MacLaughlin,~J.~A.; Doppelt,~S.~H. \emph{Science} \textbf{1981},
  \emph{211}, 590–593\relax
\mciteBstWouldAddEndPuncttrue
\mciteSetBstMidEndSepPunct{\mcitedefaultmidpunct}
{\mcitedefaultendpunct}{\mcitedefaultseppunct}\relax
\EndOfBibitem
\bibitem[Norval et~al.(2010)Norval, Bjorn, and de~Gruijl]{Norval2010}
Norval,~M.; Bjorn,~L.~O.; de~Gruijl,~F.~R. \emph{Photochem. Photobiol. Sci.}
  \textbf{2010}, \emph{9}, 11\relax
\mciteBstWouldAddEndPuncttrue
\mciteSetBstMidEndSepPunct{\mcitedefaultmidpunct}
{\mcitedefaultendpunct}{\mcitedefaultseppunct}\relax
\EndOfBibitem
\bibitem[van Dijk et~al.(2016)van Dijk, den Outer, van Kranen, and
  Slaper]{vanDijk2016}
van Dijk,~A.; den Outer,~P.; van Kranen,~H.; Slaper,~H. \emph{Photochem.
  Photobiol. Sci.} \textbf{2016}, \emph{15}, 896--909\relax
\mciteBstWouldAddEndPuncttrue
\mciteSetBstMidEndSepPunct{\mcitedefaultmidpunct}
{\mcitedefaultendpunct}{\mcitedefaultseppunct}\relax
\EndOfBibitem
\bibitem[Andreo(2015)]{andreo2015generation}
Andreo,~K. Generation of previtamin D3 from tachysterol3: A novel approach for
  producing vitamin D3 in the winter. Ph.D.\ thesis, BOSTON UNIVERSITY,
  2015\relax
\mciteBstWouldAddEndPuncttrue
\mciteSetBstMidEndSepPunct{\mcitedefaultmidpunct}
{\mcitedefaultendpunct}{\mcitedefaultseppunct}\relax
\EndOfBibitem
\bibitem[Tapavicza et~al.(2011)Tapavicza, Meyer, and Furche]{Tapavicza2011}
Tapavicza,~E.; Meyer,~A.~M.; Furche,~F. \emph{Phys. Chem. Chem. Phys.}
  \textbf{2011}, \emph{13}, 20986\relax
\mciteBstWouldAddEndPuncttrue
\mciteSetBstMidEndSepPunct{\mcitedefaultmidpunct}
{\mcitedefaultendpunct}{\mcitedefaultseppunct}\relax
\EndOfBibitem
\bibitem[Schalk et~al.(2016)Schalk, Geng, Thompson, Baluyot, Thomas, Tapavicza,
  and Hansson]{Schalk2016}
Schalk,~O.; Geng,~T.; Thompson,~T.; Baluyot,~N.; Thomas,~R.~D.; Tapavicza,~E.;
  Hansson,~T. \emph{J. Phys. Chem. A} \textbf{2016}, \emph{120}, 2320\relax
\mciteBstWouldAddEndPuncttrue
\mciteSetBstMidEndSepPunct{\mcitedefaultmidpunct}
{\mcitedefaultendpunct}{\mcitedefaultseppunct}\relax
\EndOfBibitem
\bibitem[Snyder et~al.(2016)Snyder, Curchod, and Martinez]{Snyder2016}
Snyder,~J.~W.; Curchod,~B. F.~E.; Martinez,~T.~J. \emph{J. Phys. Chem. Lett.}
  \textbf{2016}, \emph{7}, 2444--2449\relax
\mciteBstWouldAddEndPuncttrue
\mciteSetBstMidEndSepPunct{\mcitedefaultmidpunct}
{\mcitedefaultendpunct}{\mcitedefaultseppunct}\relax
\EndOfBibitem
\bibitem[Lei et~al.(2016)Lei, Wu, Zheng, Zhai, and Zhu]{lei2016photo}
Lei,~Y.; Wu,~H.; Zheng,~X.; Zhai,~G.; Zhu,~C. \emph{{J. Photochem. Photobiol.
  A: Chem.}} \textbf{2016}, \emph{317}, 39--49\relax
\mciteBstWouldAddEndPuncttrue
\mciteSetBstMidEndSepPunct{\mcitedefaultmidpunct}
{\mcitedefaultendpunct}{\mcitedefaultseppunct}\relax
\EndOfBibitem
\bibitem[Fuss et~al.(1996)Fuss, Hofer, Hering, Kompa, Lochbrunner, Schikarski,
  and Schmid]{Fuss1996}
Fuss,~W.; Hofer,~T.; Hering,~P.; Kompa,~K.~L.; Lochbrunner,~S.; Schikarski,~T.;
  Schmid,~W.~E. \emph{J. Chem. Phys.} \textbf{1996}, \emph{100}, 921\relax
\mciteBstWouldAddEndPuncttrue
\mciteSetBstMidEndSepPunct{\mcitedefaultmidpunct}
{\mcitedefaultendpunct}{\mcitedefaultseppunct}\relax
\EndOfBibitem
\bibitem[Fuss et~al.(2000)Fuss, Schmid, and Trushin]{Fuss2000}
Fuss,~W.; Schmid,~W.~E.; Trushin,~S.~A. \emph{J. Chem. Phys.} \textbf{2000},
  \emph{112}, 8347\relax
\mciteBstWouldAddEndPuncttrue
\mciteSetBstMidEndSepPunct{\mcitedefaultmidpunct}
{\mcitedefaultendpunct}{\mcitedefaultseppunct}\relax
\EndOfBibitem
\bibitem[Anderson et~al.(1999)Anderson, Shiang, and Sension]{Anderson1999}
Anderson,~N.~A.; Shiang,~J.~J.; Sension,~R.~J. \emph{J. Phys. Chem. A}
  \textbf{1999}, \emph{103}, 10730\relax
\mciteBstWouldAddEndPuncttrue
\mciteSetBstMidEndSepPunct{\mcitedefaultmidpunct}
{\mcitedefaultendpunct}{\mcitedefaultseppunct}\relax
\EndOfBibitem
\bibitem[Tang et~al.(2011)Tang, Rury, Orozco, Egendorf, Spears, and
  Sension]{Tang2011}
Tang,~K.-C.; Rury,~A.; Orozco,~M.~B.; Egendorf,~J.; Spears,~K.~G.;
  Sension,~R.~J. \emph{J. Chem. Phys.} \textbf{2011}, \emph{134}, 104503\relax
\mciteBstWouldAddEndPuncttrue
\mciteSetBstMidEndSepPunct{\mcitedefaultmidpunct}
{\mcitedefaultendpunct}{\mcitedefaultseppunct}\relax
\EndOfBibitem
\bibitem[Arruda et~al.(2013)Arruda, Peng, Smith, Spears, and
  Sension]{Arruda2013}
Arruda,~B.~C.; Peng,~J.; Smith,~B.; Spears,~K.~G.; Sension,~R.~J. \emph{J.
  Phys. Chem. B} \textbf{2013}, \emph{117}, 4696--4704\relax
\mciteBstWouldAddEndPuncttrue
\mciteSetBstMidEndSepPunct{\mcitedefaultmidpunct}
{\mcitedefaultendpunct}{\mcitedefaultseppunct}\relax
\EndOfBibitem
\bibitem[Smith et~al.(2016)Smith, Spears, and Sension]{Smith2016}
Smith,~B.~D.; Spears,~K.~G.; Sension,~R.~J. \emph{J. Phys. Chem. A}
  \textbf{2016}, \emph{120}, 6575--6581\relax
\mciteBstWouldAddEndPuncttrue
\mciteSetBstMidEndSepPunct{\mcitedefaultmidpunct}
{\mcitedefaultendpunct}{\mcitedefaultseppunct}\relax
\EndOfBibitem
\bibitem[Jacobs et~al.(1981)Jacobs, Gielen, and Havinga]{Jacobs1981}
Jacobs,~H. J.~C.; Gielen,~J. W.~J.; Havinga,~E. \emph{Tetrahedron Lett.}
  \textbf{1981}, \emph{22}, 4013\relax
\mciteBstWouldAddEndPuncttrue
\mciteSetBstMidEndSepPunct{\mcitedefaultmidpunct}
{\mcitedefaultendpunct}{\mcitedefaultseppunct}\relax
\EndOfBibitem
\bibitem[Mueller et~al.(1998)Mueller, Lochbrunner, Schmid, and
  Fuss]{Mueller1998}
Mueller,~A.~M.; Lochbrunner,~S.; Schmid,~W.~E.; Fuss,~W. \emph{Angew. Chem.
  Intl. Ed.} \textbf{1998}, \emph{37}, 505\relax
\mciteBstWouldAddEndPuncttrue
\mciteSetBstMidEndSepPunct{\mcitedefaultmidpunct}
{\mcitedefaultendpunct}{\mcitedefaultseppunct}\relax
\EndOfBibitem
\bibitem[Redwood et~al.(2013)Redwood, Bayda, and
  Saltiel]{redwood2013photoisomerization}
Redwood,~C.; Bayda,~M.; Saltiel,~J. \emph{J. Phys. Chem. Lett.} \textbf{2013},
  \emph{4}, 716--721\relax
\mciteBstWouldAddEndPuncttrue
\mciteSetBstMidEndSepPunct{\mcitedefaultmidpunct}
{\mcitedefaultendpunct}{\mcitedefaultseppunct}\relax
\EndOfBibitem
\bibitem[F~Holick(2011)]{Holick2011}
F~Holick,~M. \emph{Curr. Drug Targets} \textbf{2011}, \emph{12}, 4--18\relax
\mciteBstWouldAddEndPuncttrue
\mciteSetBstMidEndSepPunct{\mcitedefaultmidpunct}
{\mcitedefaultendpunct}{\mcitedefaultseppunct}\relax
\EndOfBibitem
\bibitem[Terenetskaya(2008)]{terenetskaya2008limitations}
Terenetskaya,~I. \emph{Theor. Exp. Chem.} \textbf{2008}, \emph{44},
  286--291\relax
\mciteBstWouldAddEndPuncttrue
\mciteSetBstMidEndSepPunct{\mcitedefaultmidpunct}
{\mcitedefaultendpunct}{\mcitedefaultseppunct}\relax
\EndOfBibitem
\bibitem[Tian et~al.(1993)Tian, Chen, Matsuoka, Wortsman, and Holick]{Tian1993}
Tian,~X.~Q.; Chen,~T.~C.; Matsuoka,~L.~Y.; Wortsman,~J.; Holick,~M.~F. \emph{J.
  Biol. Chem.} \textbf{1993}, \emph{268}, 14888\relax
\mciteBstWouldAddEndPuncttrue
\mciteSetBstMidEndSepPunct{\mcitedefaultmidpunct}
{\mcitedefaultendpunct}{\mcitedefaultseppunct}\relax
\EndOfBibitem
\bibitem[Tian and Holick(1999)Tian, and Holick]{Tian1999}
Tian,~X.~Q.; Holick,~M.~F. \emph{J. Biol. Chem.} \textbf{1999}, \emph{274},
  4174\relax
\mciteBstWouldAddEndPuncttrue
\mciteSetBstMidEndSepPunct{\mcitedefaultmidpunct}
{\mcitedefaultendpunct}{\mcitedefaultseppunct}\relax
\EndOfBibitem
\bibitem[Elliott et~al.(2000)Elliott, Ahlrichs, Hampe, and Kappes]{Elliott2000}
Elliott,~S.~D.; Ahlrichs,~R.; Hampe,~O.; Kappes,~M.~M. \emph{Phys. Chem. Chem.
  Phys.} \textbf{2000}, \emph{2}, 3415\relax
\mciteBstWouldAddEndPuncttrue
\mciteSetBstMidEndSepPunct{\mcitedefaultmidpunct}
{\mcitedefaultendpunct}{\mcitedefaultseppunct}\relax
\EndOfBibitem
\bibitem[Sugita and Okamoto(1999)Sugita, and Okamoto]{Sugita1999}
Sugita,~Y.; Okamoto,~Y. \emph{Chem. Phys. Lett.} \textbf{1999}, \emph{314},
  141\relax
\mciteBstWouldAddEndPuncttrue
\mciteSetBstMidEndSepPunct{\mcitedefaultmidpunct}
{\mcitedefaultendpunct}{\mcitedefaultseppunct}\relax
\EndOfBibitem
\bibitem[Nos{\'e}(1984)]{Nose1984}
Nos{\'e},~S. \emph{Rev. Mod. Phys.} \textbf{1984}, \emph{52}, 255\relax
\mciteBstWouldAddEndPuncttrue
\mciteSetBstMidEndSepPunct{\mcitedefaultmidpunct}
{\mcitedefaultendpunct}{\mcitedefaultseppunct}\relax
\EndOfBibitem
\bibitem[Hoover(1985)]{Hoover1985}
Hoover,~W. \emph{Phys. Rev. A} \textbf{1985}, \emph{31}, 1695\relax
\mciteBstWouldAddEndPuncttrue
\mciteSetBstMidEndSepPunct{\mcitedefaultmidpunct}
{\mcitedefaultendpunct}{\mcitedefaultseppunct}\relax
\EndOfBibitem
\bibitem[Perdew et~al.(1996)Perdew, Burke, and Ernzerhof]{Perdew1996}
Perdew,~J.; Burke,~K.; Ernzerhof,~M. \emph{Phys. Rev. Lett.} \textbf{1996},
  \emph{77}, 3865\relax
\mciteBstWouldAddEndPuncttrue
\mciteSetBstMidEndSepPunct{\mcitedefaultmidpunct}
{\mcitedefaultendpunct}{\mcitedefaultseppunct}\relax
\EndOfBibitem
\bibitem[Sch{\"a}fer et~al.(1992)Sch{\"a}fer, Horn, and Ahlrichs]{SVP}
Sch{\"a}fer,~A.; Horn,~H.; Ahlrichs,~R. \emph{J. Chem. Phys.} \textbf{1992},
  \emph{2571}, 97\relax
\mciteBstWouldAddEndPuncttrue
\mciteSetBstMidEndSepPunct{\mcitedefaultmidpunct}
{\mcitedefaultendpunct}{\mcitedefaultseppunct}\relax
\EndOfBibitem
\bibitem[Eichkorn et~al.(1997)Eichkorn, Weigend, Treutler, and
  Ahlrichs]{Eichkorn1997}
Eichkorn,~K.; Weigend,~F.; Treutler,~O.; Ahlrichs,~R. \emph{Theor. Chem. Acc.}
  \textbf{1997}, \emph{97}, 119--124\relax
\mciteBstWouldAddEndPuncttrue
\mciteSetBstMidEndSepPunct{\mcitedefaultmidpunct}
{\mcitedefaultendpunct}{\mcitedefaultseppunct}\relax
\EndOfBibitem
\bibitem[Epstein et~al.(2013)Epstein, Tapavicza, Furche, and
  Nizkorodov]{Epstein2013}
Epstein,~S.~A.; Tapavicza,~E.; Furche,~F.; Nizkorodov,~S.~A. \emph{Atmos. Chem.
  Phys.} \textbf{2013}, \emph{13}, 9461--9477\relax
\mciteBstWouldAddEndPuncttrue
\mciteSetBstMidEndSepPunct{\mcitedefaultmidpunct}
{\mcitedefaultendpunct}{\mcitedefaultseppunct}\relax
\EndOfBibitem
\bibitem[Tapavicza et~al.(2013)Tapavicza, Bellchambers, Vincent, and
  Furche]{Tapavicza2013}
Tapavicza,~E.; Bellchambers,~G.~D.; Vincent,~J.~C.; Furche,~F. \emph{Phys.
  Chem. Chem. Phys.} \textbf{2013}, \emph{15}, 18336--18348\relax
\mciteBstWouldAddEndPuncttrue
\mciteSetBstMidEndSepPunct{\mcitedefaultmidpunct}
{\mcitedefaultendpunct}{\mcitedefaultseppunct}\relax
\EndOfBibitem
\bibitem[Christiansen et~al.(1995)Christiansen, Koch, and
  J{\o}rgensen]{Christiansen95}
Christiansen,~O.; Koch,~H.; J{\o}rgensen,~P. \emph{Chem. Phys. Lett.}
  \textbf{1995}, \emph{243}, 409\relax
\mciteBstWouldAddEndPuncttrue
\mciteSetBstMidEndSepPunct{\mcitedefaultmidpunct}
{\mcitedefaultendpunct}{\mcitedefaultseppunct}\relax
\EndOfBibitem
\bibitem[H{\"a}ttig and K{\"o}hn(2002)H{\"a}ttig, and K{\"o}hn]{Hattig02}
H{\"a}ttig,~C.; K{\"o}hn,~A. \emph{J. Chem. Phys.} \textbf{2002}, \emph{117},
  6939--6951\relax
\mciteBstWouldAddEndPuncttrue
\mciteSetBstMidEndSepPunct{\mcitedefaultmidpunct}
{\mcitedefaultendpunct}{\mcitedefaultseppunct}\relax
\EndOfBibitem
\bibitem[H\"attig(2003)]{Hattig03}
H\"attig,~C. \emph{J. Chem. Phys.} \textbf{2003}, \emph{118}, 7751--7761\relax
\mciteBstWouldAddEndPuncttrue
\mciteSetBstMidEndSepPunct{\mcitedefaultmidpunct}
{\mcitedefaultendpunct}{\mcitedefaultseppunct}\relax
\EndOfBibitem
\bibitem[Tapavicza et~al.(2016)Tapavicza, Furche, and Sundholm]{Tapavicza2016}
Tapavicza,~E.; Furche,~F.; Sundholm,~D. \emph{J. Chem. Theory Comput.}
  \textbf{2016}, \emph{12}, 5058--5066\relax
\mciteBstWouldAddEndPuncttrue
\mciteSetBstMidEndSepPunct{\mcitedefaultmidpunct}
{\mcitedefaultendpunct}{\mcitedefaultseppunct}\relax
\EndOfBibitem
\bibitem[Tully(1990)]{Tully1990}
Tully,~J.~C. \emph{J. Chem. Phys.} \textbf{1990}, \emph{93}, 1061\relax
\mciteBstWouldAddEndPuncttrue
\mciteSetBstMidEndSepPunct{\mcitedefaultmidpunct}
{\mcitedefaultendpunct}{\mcitedefaultseppunct}\relax
\EndOfBibitem
\bibitem[Doltsinis and Marx(2002)Doltsinis, and Marx]{Doltsinis2002}
Doltsinis,~N.~L.; Marx,~D. \emph{Phys. Rev. Lett.} \textbf{2002}, \emph{88},
  166402\relax
\mciteBstWouldAddEndPuncttrue
\mciteSetBstMidEndSepPunct{\mcitedefaultmidpunct}
{\mcitedefaultendpunct}{\mcitedefaultseppunct}\relax
\EndOfBibitem
\bibitem[Barbatti et~al.(2007)Barbatti, Granucci, Persico, Ruckenbauer, Vazdar,
  Eckert-Maksic, and Lischka]{Barbatti2007}
Barbatti,~M.; Granucci,~G.; Persico,~M.; Ruckenbauer,~M.; Vazdar,~M.;
  Eckert-Maksic,~M.; Lischka,~H. \emph{{J. Photochem. Photobiol. A: Chem.}}
  \textbf{2007}, \emph{190}, 228\relax
\mciteBstWouldAddEndPuncttrue
\mciteSetBstMidEndSepPunct{\mcitedefaultmidpunct}
{\mcitedefaultendpunct}{\mcitedefaultseppunct}\relax
\EndOfBibitem
\bibitem[Tapavicza et~al.(2009)Tapavicza, Tavernelli, and
  Rothlisberger]{Tapavicza2009}
Tapavicza,~E.; Tavernelli,~I.; Rothlisberger,~U. \emph{J. Phys. Chem. A}
  \textbf{2009}, \emph{113}, 9595\relax
\mciteBstWouldAddEndPuncttrue
\mciteSetBstMidEndSepPunct{\mcitedefaultmidpunct}
{\mcitedefaultendpunct}{\mcitedefaultseppunct}\relax
\EndOfBibitem
\bibitem[Barbatti(2011)]{barbatti2011nonadiabatic}
Barbatti,~M. \emph{Wiley Interdiscip. Rev. Comput. Mol. Sci.} \textbf{2011},
  \emph{1}, 620--633\relax
\mciteBstWouldAddEndPuncttrue
\mciteSetBstMidEndSepPunct{\mcitedefaultmidpunct}
{\mcitedefaultendpunct}{\mcitedefaultseppunct}\relax
\EndOfBibitem
\bibitem[Nelson et~al.(2011)Nelson, Fernandez-Alberti, Chernyak, Roitberg, and
  Tretiak]{Nelson2011}
Nelson,~T.; Fernandez-Alberti,~S.; Chernyak,~V.; Roitberg,~A.~E.; Tretiak,~S.
  \emph{J. Phys. Chem. B} \textbf{2011}, \emph{115}, 5402--5414\relax
\mciteBstWouldAddEndPuncttrue
\mciteSetBstMidEndSepPunct{\mcitedefaultmidpunct}
{\mcitedefaultendpunct}{\mcitedefaultseppunct}\relax
\EndOfBibitem
\bibitem[Mitri\'c et~al.(2011)Mitri\'c, Petersen, Wohlgemuth, Werner,
  Bona\v{c}i\'{c}-Kouteck\'y, W\"oste, and Jortner]{Mitric2011a}
Mitri\'c,~R.; Petersen,~J.; Wohlgemuth,~M.; Werner,~U.;
  Bona\v{c}i\'{c}-Kouteck\'y,~V.; W\"oste,~L.; Jortner,~J. \emph{J. Phys. Chem.
  A} \textbf{2011}, \emph{115}, 3755\relax
\mciteBstWouldAddEndPuncttrue
\mciteSetBstMidEndSepPunct{\mcitedefaultmidpunct}
{\mcitedefaultendpunct}{\mcitedefaultseppunct}\relax
\EndOfBibitem
\bibitem[Yu et~al.(2014)Yu, Xu, Lei, Zhu, and Wen]{yu2014trajectory}
Yu,~L.; Xu,~C.; Lei,~Y.; Zhu,~C.; Wen,~Z. \emph{Phys. Chem. Chem. Phys.}
  \textbf{2014}, \emph{16}, 25883--25895\relax
\mciteBstWouldAddEndPuncttrue
\mciteSetBstMidEndSepPunct{\mcitedefaultmidpunct}
{\mcitedefaultendpunct}{\mcitedefaultseppunct}\relax
\EndOfBibitem
\bibitem[Curchod et~al.(2013)Curchod, Rothlisberger, and
  Tavernelli]{curchod2013trajectory}
Curchod,~B.~F.; Rothlisberger,~U.; Tavernelli,~I. \emph{Chem. Phys. Chem.}
  \textbf{2013}, \emph{14}, 1314--1340\relax
\mciteBstWouldAddEndPuncttrue
\mciteSetBstMidEndSepPunct{\mcitedefaultmidpunct}
{\mcitedefaultendpunct}{\mcitedefaultseppunct}\relax
\EndOfBibitem
\bibitem[Granucci and Persico(2007)Granucci, and Persico]{Granucci2007}
Granucci,~G.; Persico,~M. \emph{J. Chem. Phys.} \textbf{2007}, \emph{126},
  134114\relax
\mciteBstWouldAddEndPuncttrue
\mciteSetBstMidEndSepPunct{\mcitedefaultmidpunct}
{\mcitedefaultendpunct}{\mcitedefaultseppunct}\relax
\EndOfBibitem
\bibitem[Domcke et~al.(2004)Domcke, Yarkony, and K{\"o}ppel]{Domcke2004conical}
Domcke,~W.; Yarkony,~D.; K{\"o}ppel,~H. \emph{Conical intersections: electronic
  structure, dynamics \& spectroscopy}; Advanced series in physical chemistry;
  World Scientific, 2004\relax
\mciteBstWouldAddEndPuncttrue
\mciteSetBstMidEndSepPunct{\mcitedefaultmidpunct}
{\mcitedefaultendpunct}{\mcitedefaultseppunct}\relax
\EndOfBibitem
\bibitem[Herman(1982)]{herman2949}
Herman,~M.~F. \emph{J. Chem. Phys.} \textbf{1982}, \emph{76}, 2949\relax
\mciteBstWouldAddEndPuncttrue
\mciteSetBstMidEndSepPunct{\mcitedefaultmidpunct}
{\mcitedefaultendpunct}{\mcitedefaultseppunct}\relax
\EndOfBibitem
\bibitem[Herman(1983)]{herman2771}
Herman,~M.~F. \emph{J. Chem. Phys.} \textbf{1983}, \emph{79}, 2771\relax
\mciteBstWouldAddEndPuncttrue
\mciteSetBstMidEndSepPunct{\mcitedefaultmidpunct}
{\mcitedefaultendpunct}{\mcitedefaultseppunct}\relax
\EndOfBibitem
\bibitem[Vincent et~al.(2016)Vincent, Muuronen, Pearce, Mohanam, Tapavicza, and
  Furche]{Vincent2016}
Vincent,~J.~C.; Muuronen,~M.; Pearce,~K.~C.; Mohanam,~L.~N.; Tapavicza,~E.;
  Furche,~F. \emph{J. Phys. Chem. Lett.} \textbf{2016}, \emph{7},
  4185--4190\relax
\mciteBstWouldAddEndPuncttrue
\mciteSetBstMidEndSepPunct{\mcitedefaultmidpunct}
{\mcitedefaultendpunct}{\mcitedefaultseppunct}\relax
\EndOfBibitem
\bibitem[Furche and Ahlrichs(2002)Furche, and Ahlrichs]{Furche2002}
Furche,~F.; Ahlrichs,~R. \emph{J. Chem. Phys.} \textbf{2002}, \emph{117},
  7433\relax
\mciteBstWouldAddEndPuncttrue
\mciteSetBstMidEndSepPunct{\mcitedefaultmidpunct}
{\mcitedefaultendpunct}{\mcitedefaultseppunct}\relax
\EndOfBibitem
\bibitem[Chernyak and Mukamel(2000)Chernyak, and Mukamel]{Chernyak2000}
Chernyak,~V.; Mukamel,~S. \emph{J. Chem. Phys.} \textbf{2000}, \emph{112},
  3572\relax
\mciteBstWouldAddEndPuncttrue
\mciteSetBstMidEndSepPunct{\mcitedefaultmidpunct}
{\mcitedefaultendpunct}{\mcitedefaultseppunct}\relax
\EndOfBibitem
\bibitem[Baer(2002)]{Baer2002}
Baer,~R. \emph{Chem. Phys. Lett.} \textbf{2002}, \emph{364}, 75\relax
\mciteBstWouldAddEndPuncttrue
\mciteSetBstMidEndSepPunct{\mcitedefaultmidpunct}
{\mcitedefaultendpunct}{\mcitedefaultseppunct}\relax
\EndOfBibitem
\bibitem[Tavernelli et~al.(2009)Tavernelli, Tapavicza, and
  Rothlisberger]{Tavernelli2009}
Tavernelli,~I.; Tapavicza,~E.; Rothlisberger,~U. \emph{J. Chem. Phys.}
  \textbf{2009}, \emph{130}, 124107\relax
\mciteBstWouldAddEndPuncttrue
\mciteSetBstMidEndSepPunct{\mcitedefaultmidpunct}
{\mcitedefaultendpunct}{\mcitedefaultseppunct}\relax
\EndOfBibitem
\bibitem[Tavernelli et~al.(2009)Tavernelli, Tapavicza, and
  Rothlisberger]{Tavernelli2009a}
Tavernelli,~I.; Tapavicza,~E.; Rothlisberger,~U. \emph{J. Mol. Struct.
  THEOCHEM} \textbf{2009}, \emph{914}, 22\relax
\mciteBstWouldAddEndPuncttrue
\mciteSetBstMidEndSepPunct{\mcitedefaultmidpunct}
{\mcitedefaultendpunct}{\mcitedefaultseppunct}\relax
\EndOfBibitem
\bibitem[Send and Furche(2010)Send, and Furche]{Send2010}
Send,~R.; Furche,~F. \emph{J. Phys. Chem.} \textbf{2010}, \emph{132},
  044107\relax
\mciteBstWouldAddEndPuncttrue
\mciteSetBstMidEndSepPunct{\mcitedefaultmidpunct}
{\mcitedefaultendpunct}{\mcitedefaultseppunct}\relax
\EndOfBibitem
\bibitem[Ou et~al.(2015)Ou, Bellchambers, Furche, and
  Subotnik]{Ou2015_nac_between_excited_states}
Ou,~Q.; Bellchambers,~G.~D.; Furche,~F.; Subotnik,~J.~E. \emph{J. Chem. Phys.}
  \textbf{2015}, \emph{142}, 064114\relax
\mciteBstWouldAddEndPuncttrue
\mciteSetBstMidEndSepPunct{\mcitedefaultmidpunct}
{\mcitedefaultendpunct}{\mcitedefaultseppunct}\relax
\EndOfBibitem
\bibitem[Tapavicza et~al.(2007)Tapavicza, Tavernelli, and
  Rothlisberger]{Tapavicza2007}
Tapavicza,~E.; Tavernelli,~I.; Rothlisberger,~U. \emph{Phys. Rev. Lett.}
  \textbf{2007}, \emph{98}, 023001\relax
\mciteBstWouldAddEndPuncttrue
\mciteSetBstMidEndSepPunct{\mcitedefaultmidpunct}
{\mcitedefaultendpunct}{\mcitedefaultseppunct}\relax
\EndOfBibitem
\bibitem[Send et~al.(2011)Send, K\"uhn, and Furche]{Send2011}
Send,~R.; K\"uhn,~M.; Furche,~F. \emph{J. Chem. Theory Comput.} \textbf{2011},
  \emph{7}, 2376--2386\relax
\mciteBstWouldAddEndPuncttrue
\mciteSetBstMidEndSepPunct{\mcitedefaultmidpunct}
{\mcitedefaultendpunct}{\mcitedefaultseppunct}\relax
\EndOfBibitem
\bibitem[Dreuw et~al.(2003)Dreuw, Weisman, and Head-Gordon]{Dreuw2003}
Dreuw,~A.; Weisman,~J.; Head-Gordon,~M. \emph{J. Chem. Phys.} \textbf{2003},
  \emph{119}, 2943\relax
\mciteBstWouldAddEndPuncttrue
\mciteSetBstMidEndSepPunct{\mcitedefaultmidpunct}
{\mcitedefaultendpunct}{\mcitedefaultseppunct}\relax
\EndOfBibitem
\bibitem[Levine et~al.(2006)Levine, Ko, Quenneville, and Martinez]{Levine2006}
Levine,~B.~G.; Ko,~C.; Quenneville,~J.; Martinez,~T.~J. \emph{Mol. Phys.}
  \textbf{2006}, \emph{104}, 1039\relax
\mciteBstWouldAddEndPuncttrue
\mciteSetBstMidEndSepPunct{\mcitedefaultmidpunct}
{\mcitedefaultendpunct}{\mcitedefaultseppunct}\relax
\EndOfBibitem
\bibitem[Christiansen et~al.(1996)Christiansen, Koch, J{\o}rgensen, and
  Helgaker]{Christiansen1996}
Christiansen,~O.; Koch,~H.; J{\o}rgensen,~P.; Helgaker,~T. \emph{Chem. Phys.
  Lett.} \textbf{1996}, \emph{263}, 530\relax
\mciteBstWouldAddEndPuncttrue
\mciteSetBstMidEndSepPunct{\mcitedefaultmidpunct}
{\mcitedefaultendpunct}{\mcitedefaultseppunct}\relax
\EndOfBibitem
\bibitem[Christiansen et~al.(1996)Christiansen, Koch, Halkier, Jorgensen,
  Helgaker, and de~Meras]{Christiansen1996a}
Christiansen,~O.; Koch,~H.; Halkier,~A.; Jorgensen,~P.; Helgaker,~T.;
  de~Meras,~A.~S. \emph{J. Chem. Phys.} \textbf{1996}, \emph{105}, 6921\relax
\mciteBstWouldAddEndPuncttrue
\mciteSetBstMidEndSepPunct{\mcitedefaultmidpunct}
{\mcitedefaultendpunct}{\mcitedefaultseppunct}\relax
\EndOfBibitem
\bibitem[Tapavicza et~al.(2008)Tapavicza, Tavernelli, Rothlisberger, Filippi,
  and Casida]{Tapavicza2008}
Tapavicza,~E.; Tavernelli,~I.; Rothlisberger,~U.; Filippi,~C.; Casida,~M.~E.
  \emph{J. Chem. Phys.} \textbf{2008}, \emph{129}, 124108\relax
\mciteBstWouldAddEndPuncttrue
\mciteSetBstMidEndSepPunct{\mcitedefaultmidpunct}
{\mcitedefaultendpunct}{\mcitedefaultseppunct}\relax
\EndOfBibitem
\bibitem[Hirata and Head-Gordon(1999)Hirata, and Head-Gordon]{Hirata1999}
Hirata,~S.; Head-Gordon,~M. \emph{Chem. Phys. Lett.} \textbf{1999}, \emph{314},
  291\relax
\mciteBstWouldAddEndPuncttrue
\mciteSetBstMidEndSepPunct{\mcitedefaultmidpunct}
{\mcitedefaultendpunct}{\mcitedefaultseppunct}\relax
\EndOfBibitem
\bibitem[TUR()]{TURBOMOLE}
{TURBOMOLE V6.3}, {TURBOMOLE GmbH}, {K}arlsruhe, 2011; available from \\ {\tt
  http://www.turbomole.com}.\relax
\mciteBstWouldAddEndPunctfalse
\mciteSetBstMidEndSepPunct{\mcitedefaultmidpunct}
{}{\mcitedefaultseppunct}\relax
\EndOfBibitem
\bibitem[H{\"a}ser and Ahlrichs(1989)H{\"a}ser, and Ahlrichs]{Haeser1989}
H{\"a}ser,~M.; Ahlrichs,~R. \emph{J. Comput. Chem.} \textbf{1989}, \emph{10},
  104\relax
\mciteBstWouldAddEndPuncttrue
\mciteSetBstMidEndSepPunct{\mcitedefaultmidpunct}
{\mcitedefaultendpunct}{\mcitedefaultseppunct}\relax
\EndOfBibitem
\bibitem[Furche(2001)]{Furche2001}
Furche,~F. \emph{J. Chem. Phys.} \textbf{2001}, \emph{114}, 5982--5992\relax
\mciteBstWouldAddEndPuncttrue
\mciteSetBstMidEndSepPunct{\mcitedefaultmidpunct}
{\mcitedefaultendpunct}{\mcitedefaultseppunct}\relax
\EndOfBibitem
\bibitem[Dmitrenko and Reischl(1997)Dmitrenko, and Reischl]{Dmitrenko1997}
Dmitrenko,~O.; Reischl,~W. \emph{Res. Chem. Intermediat.} \textbf{1997},
  \emph{23}, 691--702\relax
\mciteBstWouldAddEndPuncttrue
\mciteSetBstMidEndSepPunct{\mcitedefaultmidpunct}
{\mcitedefaultendpunct}{\mcitedefaultseppunct}\relax
\EndOfBibitem
\bibitem[Ramakrishnan et~al.(2015)Ramakrishnan, Hartmann, Tapavicza, and von
  Lilienfeld]{Ramakrishnan2015}
Ramakrishnan,~R.; Hartmann,~M.; Tapavicza,~E.; von Lilienfeld,~O.~A. \emph{J.
  Chem. Phys.} \textbf{2015}, \emph{143}, 084111\relax
\mciteBstWouldAddEndPuncttrue
\mciteSetBstMidEndSepPunct{\mcitedefaultmidpunct}
{\mcitedefaultendpunct}{\mcitedefaultseppunct}\relax
\EndOfBibitem
\bibitem[Saltiel et~al.(2003)Saltiel, Cires, and Turek]{Saltiel2003}
Saltiel,~J.; Cires,~L.; Turek,~A.~M. \emph{J. Am. Chem. Soc.} \textbf{2003},
  \emph{125}, 2866\relax
\mciteBstWouldAddEndPuncttrue
\mciteSetBstMidEndSepPunct{\mcitedefaultmidpunct}
{\mcitedefaultendpunct}{\mcitedefaultseppunct}\relax
\EndOfBibitem
\bibitem[Green et~al.(1974)Green, Sawada, and Shettle]{green1974middle}
Green,~A.; Sawada,~T.; Shettle,~E. \emph{Photochem. Photobiol.} \textbf{1974},
  \emph{19}, 251--259\relax
\mciteBstWouldAddEndPuncttrue
\mciteSetBstMidEndSepPunct{\mcitedefaultmidpunct}
{\mcitedefaultendpunct}{\mcitedefaultseppunct}\relax
\EndOfBibitem
\bibitem[Boomsma et~al.(1975)Boomsma, Jacobs, Havinga, and van~der
  Gen]{Boomsma1975}
Boomsma,~F.; Jacobs,~H.; Havinga,~E.; van~der Gen,~A. \emph{Tetrahedron Lett.}
  \textbf{1975}, \emph{16}, 427\relax
\mciteBstWouldAddEndPuncttrue
\mciteSetBstMidEndSepPunct{\mcitedefaultmidpunct}
{\mcitedefaultendpunct}{\mcitedefaultseppunct}\relax
\EndOfBibitem
\bibitem[Boomsma et~al.(1977)Boomsma, Jacobs, Havinga, and Van~der
  Gen]{Boomsma1977overirradiation}
Boomsma,~F.; Jacobs,~H.; Havinga,~E.; Van~der Gen,~A. \emph{Recl. Trav. Chim.
  Pays-Bas} \textbf{1977}, \emph{96}, 104--112\relax
\mciteBstWouldAddEndPuncttrue
\mciteSetBstMidEndSepPunct{\mcitedefaultmidpunct}
{\mcitedefaultendpunct}{\mcitedefaultseppunct}\relax
\EndOfBibitem
\bibitem[Maessen et~al.(1983)Maessen, Jacobs, Cornelisse, and
  Havinga]{maessen1983photochemistry}
Maessen,~P.~A.; Jacobs,~H.~J.; Cornelisse,~J.; Havinga,~E. \emph{Angew. Chem.
  Intl. Ed.} \textbf{1983}, \emph{22}, 718--719\relax
\mciteBstWouldAddEndPuncttrue
\mciteSetBstMidEndSepPunct{\mcitedefaultmidpunct}
{\mcitedefaultendpunct}{\mcitedefaultseppunct}\relax
\EndOfBibitem
\bibitem[Garavelli et~al.(1997)Garavelli, Celani, Bernardi, Robb, and
  Olivucci]{Garavelli1997_trans-hexatriene}
Garavelli,~M.; Celani,~P.; Bernardi,~F.; Robb,~M.~A.; Olivucci,~M. \emph{J. Am.
  Chem. Soc.} \textbf{1997}, \emph{119}, 11487--11494\relax
\mciteBstWouldAddEndPuncttrue
\mciteSetBstMidEndSepPunct{\mcitedefaultmidpunct}
{\mcitedefaultendpunct}{\mcitedefaultseppunct}\relax
\EndOfBibitem
\bibitem[Anderson et~al.(2000)Anderson, III, Murnane, Kapteyn, and
  Sension]{Anderson2000_cis_trans_hexatriene}
Anderson,~N.~A.; III,~C. G.~D.; Murnane,~M.~M.; Kapteyn,~H.~C.; Sension,~R.~J.
  \emph{Chem. Phys. Lett.} \textbf{2000}, \emph{323}, 365 -- 371\relax
\mciteBstWouldAddEndPuncttrue
\mciteSetBstMidEndSepPunct{\mcitedefaultmidpunct}
{\mcitedefaultendpunct}{\mcitedefaultseppunct}\relax
\EndOfBibitem
\bibitem[Dmitrenko et~al.(1999)Dmitrenko, Vivian, Reischl, and
  Frederick]{Dmitrenko1999}
Dmitrenko,~O.; Vivian,~J.~T.; Reischl,~W.; Frederick,~J.~H. \emph{J. Mol.
  Struct. THEOCHEM} \textbf{1999}, \emph{467}, 195 -- 210\relax
\mciteBstWouldAddEndPuncttrue
\mciteSetBstMidEndSepPunct{\mcitedefaultmidpunct}
{\mcitedefaultendpunct}{\mcitedefaultseppunct}\relax
\EndOfBibitem
\bibitem[Tuckerman et~al.(1997)Tuckerman, Marx, Klein, and
  Parrinello]{Tuckerman1997}
Tuckerman,~M.~E.; Marx,~D.; Klein,~M.~L.; Parrinello,~M. \emph{Science}
  \textbf{1997}, \emph{275}, 817--820\relax
\mciteBstWouldAddEndPuncttrue
\mciteSetBstMidEndSepPunct{\mcitedefaultmidpunct}
{\mcitedefaultendpunct}{\mcitedefaultseppunct}\relax
\EndOfBibitem
\bibitem[Zimmermann and Van{\'i}{\v{c}}ek(2010)Zimmermann, and
  Van{\'i}{\v{c}}ek]{Zimmermann2010}
Zimmermann,~T.; Van{\'i}{\v{c}}ek,~J. \emph{J. Mol. Model.} \textbf{2010},
  \emph{16}, 1779--1787\relax
\mciteBstWouldAddEndPuncttrue
\mciteSetBstMidEndSepPunct{\mcitedefaultmidpunct}
{\mcitedefaultendpunct}{\mcitedefaultseppunct}\relax
\EndOfBibitem
\bibitem[Havinga(1973)]{Havinga1973}
Havinga,~E. \emph{Cell. Mol. Life Sci.} \textbf{1973}, \emph{29}, 1181\relax
\mciteBstWouldAddEndPuncttrue
\mciteSetBstMidEndSepPunct{\mcitedefaultmidpunct}
{\mcitedefaultendpunct}{\mcitedefaultseppunct}\relax
\EndOfBibitem
\bibitem[ast()]{astm}
ASTM G173-03(2012), Standard Tables for Reference Solar Spectral Irradiances:
  Direct Normal and Hemispherical on 37° Tilted Surface, ASTM International,
  West Conshohocken, PA, 2012. \url{ www.astm.org}, Accessed: 2016-10-24\relax
\mciteBstWouldAddEndPuncttrue
\mciteSetBstMidEndSepPunct{\mcitedefaultmidpunct}
{\mcitedefaultendpunct}{\mcitedefaultseppunct}\relax
\EndOfBibitem
\bibitem[Khintchine(1929)]{Khintchine1929}
Khintchine,~A. \emph{Comtes rendus de l’Acad{\'e}mie des sciences}
  \textbf{1929}, \emph{188}, 477--479\relax
\mciteBstWouldAddEndPuncttrue
\mciteSetBstMidEndSepPunct{\mcitedefaultmidpunct}
{\mcitedefaultendpunct}{\mcitedefaultseppunct}\relax
\EndOfBibitem
\bibitem[Grinstead and Snell(2012)Grinstead, and
  Snell]{grinstead2012introduction}
Grinstead,~C.~M.; Snell,~J.~L. \emph{Introduction to probability}; American
  Mathematical Soc., 2012\relax
\mciteBstWouldAddEndPuncttrue
\mciteSetBstMidEndSepPunct{\mcitedefaultmidpunct}
{\mcitedefaultendpunct}{\mcitedefaultseppunct}\relax
\EndOfBibitem
\bibitem[sol()]{solar_position}
Solar Position Calculator - U.S. Department of Commerce, National Oceanic \&
  Atmospheric Administration, NOAA Research.
  \url{http://www.esrl.noaa.gov/gmd/grad/solcalc/azel.html}, Accessed:
  2016-10-24\relax
\mciteBstWouldAddEndPuncttrue
\mciteSetBstMidEndSepPunct{\mcitedefaultmidpunct}
{\mcitedefaultendpunct}{\mcitedefaultseppunct}\relax
\EndOfBibitem
\end{mcitethebibliography}
\end{document}